# On the Arbitrary Choice Regarding Which Inertial Reference Frame is "Stationary" and Which is "Moving" in the Special Theory of Relativity

Douglas M. Snyder
Los Angeles, CA

The relativity of simultaneity is central to the special theory, and it is the basis for the other results of the special theory. It is the relativity of simultaneity that fundamentally distinguishes the special theory from Newtonian mechanics and the kinematics underlying it. What generally is not appreciated is that Einstein's *argument* on the relativity of simultaneity itself is the first result and that the argument itself is reflected in the structure and functioning of the physical world. The arbitrary nature of the decision regarding the particular inertial reference frame from which Einstein's argument on the relativity of simultaneity begins is discussed, and it is this arbitrary, or freely made, decision that is the basis for the significance of the argument of the relativity of simultaneity itself on the structure and functioning of the physical world. Moreover, the arbitrary choice as to the direction in which the relativity of simultaneity is argued indicates that there is a specific cognitive effect on the functioning of the physical world because the particular format of the argument on the relativity of simultaneity used is freely chosen without physical or mental constraints, as none are indicated in the special theory.

The paper approaches the role of the relativity of simultaneity in the special theory through a gedankenexperiment known as the twin paradox and, more generally, in terms of the relation between temporal durations of an occurrence and the spatial lengths of physical existents in inertial reference frames in uniform translational motion relative to one another. The argument concerning the relativity of simultaneity leads to an interesting conundrum regarding light, a conundrum that serves to make more explicit the importance of the argument on the relativity of simultaneity to the structure and functioning of the physical world in the special theory. The general lack of two-way empirical tests in the special theory (i.e., tests in which each of the two inertial reference frames is considered the "stationary" reference frame, the reference frame in which the argument on the relativity of simultaneity begins, in a scenario) is discussed. Two examples of other kinds of tests which have been conducted are discussed and contrasted with a proposed two-way test.

# On the Arbitrary Choice

*1.    The Twin Paradox*

A gedankenexperiment known as the twin paradox has attracted much attention since Einstein proposed the special theory of relativity. Essentially, in the twin paradox one of a pair of twins, who are both originally on earth, travels in a spaceship away from the earth with constant velocity in a rectilinear manner and without rotation. After traveling away from the earth, the spaceship makes an abrupt $180^o$ turn and returns to earth. The question asked in the paradox is: Which twin is younger and which is older upon the return of the spaceship to earth? According to a popular phrase conveying a fundamental result of the special theory of relativity, moving clocks run slower. Thus, as each of the twins considers for the most part himself or herself at rest and his or her sibling moving in a uniform, rectilinear manner relative to him or her, each sibling would apparently expect to find the other sibling younger than he or she when the spaceship returns to earth. The "moving" sibling finds himself or herself aging in accordance with the time measured by the "moving" clocks, and thus this sibling would age slower. In fact, according to Feynman, Leighton, and Sands, the biological aging processes in the "moving" sibling or the "stationary" sibling themselves could be considered clocks at rest in the respective inertial reference frames in which each of the siblings is at rest.

The underlying principle in this apparent paradox is that for two inertial reference frames in uniform translational motion relative to one another, it is arbitrary in the special theory which inertial frame is considered the "stationary" frame and which the "moving" frame for deducing the Lorentz coordinate transformation equations for these inertial reference frames. (As the Lorentz coordinate transformation equations are dependent on the relativity of simultaneity, as will be shown, "stationary" essentially refers to the inertial reference frame where the argument concerning the relativity of simultaneity begins, where simultaneity is first established and time delineated for an inertial reference frame. "Moving" essentially refers to the other inertial reference frame where the time of the "stationary" reference frame is introduced to determine whether the criterion of simultaneity in the "moving" frame is met.) As noted, in the special theory, the Lorentz coordinate transformation equations depend fundamentally on the relativity of simultaneity for inertial reference frames in uniform translational motion relative to one another. And the relativity of simultaneity for these inertial reference frames depends on two major factors: 1) the use of the invariant and finite velocity of light in all inertial reference frames in the development of simultaneity and time in an inertial

# On the Arbitrary Choice

reference frame; and 2) the principle of relativity that the laws of physics are the same in all inertial reference frames. Both factors prevent an observer at rest in one of two inertial reference frames moving in a uniform translational manner relative to each other from having a privileged view concerning the functioning of the physical world. Neither observer can rely on the motion of light to determine which inertial frame is actually moving and which stationary. And neither observer can rely on the relative velocity of the two inertial reference frames to help in this determination.

As concerns the twin paradox, the attempted resolution most widely held depends on the point that only the twin in the spaceship experiences the acceleration resulting from the spaceship turning around in order to place it on a course toward earth. Thus, there would appear to be some "objective" basis for determining that the twin on the spaceship was really moving.[(1)] As Feynman, Leighton, and Sands wrote concerning the twin paradox:

> So the way to state the rule is to say that the man who has felt the accelerations, who has seen things fall against the walls, and so on, is the one who would be the younger; that is the difference between them in an "absolute" sense, and it is certainly correct.[(2)]

Various other explanations have been suggested over the years from both the special theory as well as the general theory of relativity to resolve the paradox.[1]

## 2. *A Different Gedankenexperiment*

Whether the attempted resolutions in fact resolve the paradox, the concern here is with a more basic scenario than the twin paradox: it is the circumstance where the spaceship continues to move away from the earth. That is, the scenario here is the spaceship does not turn around abruptly and return to earth. Essentially, this scenario involves two inertial reference frames that continue to move in a uniform translational manner relative to one another. With regard to this circumstance, one cannot, for example, attempt to rely on the acceleration of one of the frames (i.e., the one "really" moving) to distinguish the two reference frames and thus to decide which clock in its respective inertial reference frame runs slower.

In the scenario of concern in this paper, what is the relation between temporal durations measured by the twins if they keep moving away from one another in a uniform translational manner? Is the twin on earth aging slower, or is the twin on the spaceship aging slower? It might seem that such questions are like the position of the critic in Feynman, Leighton, and Sands' discussion of the twin paradox. They wrote:

> [The twin paradox] is called a "paradox" only by the people who believe that the [special] principle of relativity means that *all motion* is relative; they say, "Heh, heh, heh from the point of view of Paul [the twin on the spaceship], can't we say that *Peter* was moving and should therefore appear to age more slowly? By symmetry, the only possible result is that both should be the same age when they meet."[3]

In the scenario where the twin in the spaceship does not turn around and return to earth, there is only uniform, translational motion. In the special theory, this kind of motion is clearly relative. It will be shown that the answer as to which inertial reference frame's clocks run "slower" and which "faster" depends fundamentally on the relationship between the inertial reference frames in uniform translational motion in the argument on the relativity of simultaneity in the special theory. The logical relationship is the key because whatever the particular relationship between the time measured by clocks at rest in their respective inertial reference frames in uniform translational motion relative to one another is depends on the relativity of simultaneity.

More specifically, the designation of one of the clocks as at rest in a "stationary" inertial reference frame (A) and the other clock at rest in a "moving" inertial reference frame (B) depends on the corresponding designations of A as "stationary" in the argument on the relativity of simultaneity and B as "moving" in this argument. There is nothing physical indicated in the special theory that keeps the argument concerning the relativity of simultaneity from being argued with B as the "stationary" reference frame and A as the "moving" reference frame. Indeed the relativity of simultaneity can be argued in this manner and the theoretical integrity of the special theory depends on it. If it were not possible to argue the relativity of simultaneity in this latter fashion, there would be a preferred inertial reference frame, a violation of the fundamental tenet of the special theory that inertial reference frames in uniform translational motion relative to one another are equivalent with regard to the description of physical phenomena.

These same points hold for the relation between temporal durations in inertial reference frames in uniform translational motion relative to one another. This last result should not be surprising as the relationship between temporal durations depends for its explanation on the argument on the relativity of simultaneity. One implication of this dependence is, for example, that in the case of the twins the notion of one twin aging slower or faster than the other occurs only in the context of its particular relationship with the other twin. The context of this relationship takes the form of a logical argument that begins with the consideration of one of the inertial reference frames in which light, having a finite and invariant velocity, is used to determine simultaneity in this reference frame (called the "stationary" frame). This same light, specifically its velocity, is then juxtaposed with the uniform translational velocity of the inertial reference frames relative to one another in order to determine whether the criterion for simultaneity in the other inertial reference frame (called the "moving" frame) is met. (Even though observers at rest in the "moving" frame measure the invariant velocity for this light as do observers in the "stationary" frame, the juxtaposition of the invariant velocity of light in the "stationary" frame with the uniform translational velocity of the inertial reference frames relative to one another is at the center of the relativity of simultaneity.)[2]

As noted, one part of the answer as to which inertial frame's clocks run "slower" and which "faster" is theoretical in nature in that it involves the logical structure of the argument demonstrating the relativity of simultaneity. The other part of the answer appears to be concrete in nature. It depends on the fact that an observer always considers himself at rest in his inertial reference frame and considers his reference frame at rest as well. Thus, an observer in an inertial reference frame always sees himself in the "stationary" inertial reference frame. If he did not see himself as such, if he knew that he was moving, the argument concerning the relativity of simultaneity could not be argued with this observer as the "stationary" observer. We would then have absolute motion, a violation of the fundamental tenet of the special theory that for observers at rest in inertial reference frames, uniform translational motion is relative. Thus, what at first appears a concrete indication of the "stationary" reference frame in that it is experienced is in part theoretical in nature. It is in part theoretical because the observer's concrete experience of being at rest in his inertial reference frame, and seeing his frame as "stationary," while obtaining empirical results consistent with his being in the "stationary" reference frame which are in agreement with the special theory, depends for its explanation on the argument demonstrating the relativity of simultaneity.

The relationship between temporal durations in inertial reference frames in uniform translational motion relative to one another has been addressed briefly. A similar question concerning the spatial length of some physical existent may be asked as well. Specifically, if observers at rest in their respective inertial reference frames each measure identically constructed rods at rest in the other observers' inertial reference frame, each observer will measure the rod in the other frame, the "moving" frame, to be shorter than the rod in their own frame in which it is at rest. Thus, who is really measuring the shorter rod? In the next two sections, these relationships are discussed in greater detail and pictorially depicted.

*3.    Further Comments on Temporal Relations Between Inertial Reference Frames*

Figure 1 is a Minkowski diagram of two inertial reference frames, W and W', each of one spatial dimension, moving in a uniform translational manner relative to one another along their respective spatial axes and in which the the description of the relative motion begins with the spatial and temporal origins of W and W' corresponding to one another. In the Minkowski diagram, this correspondence of the spatial and temporal origins is represented by the spatial and temporal origins of the coordinate schemes for W and W' overlapping. For inertial reference frame W, axis x and t are the space and time axes respectively. For inertial reference frame W', x' and t' are the space and time axes respectively.

The general expressions relating temporal durations in inertial reference frames in uniform translational motion relative to one another are:

$$\Delta t' = \Delta t/(1 - v^2/c^2)^{1/2} \quad (1)$$

where W' is considered the "stationary" inertial reference frame and W the "moving" inertial reference frame, and:

$$\Delta t = \Delta t'/(1 - v^2/c^2)^{1/2} \quad (2)$$

where W is considered the "stationary" inertial reference frame and W' the "moving" inertial reference frame. $\Delta t = t_2 - t_1$, $\Delta t' = t'_2 - t'_1$, v is the magnitude of the uniform translational velocity of W and W' relative to one another, and c is the invariant velocity of light in inertial reference frames. As will be shown in "Spatiotemporal Relations and the Lorentz Transformation Equations," equations 1 and 2 are derived from the Lorentz coordinate transformation equations.

First, consider the scenario where an observer at rest in W' considers the time measured by a clock at rest in W. With regard to W', the clock at rest in W is moving with a uniform, translational velocity, -v. This particular

# On the Arbitrary Choice

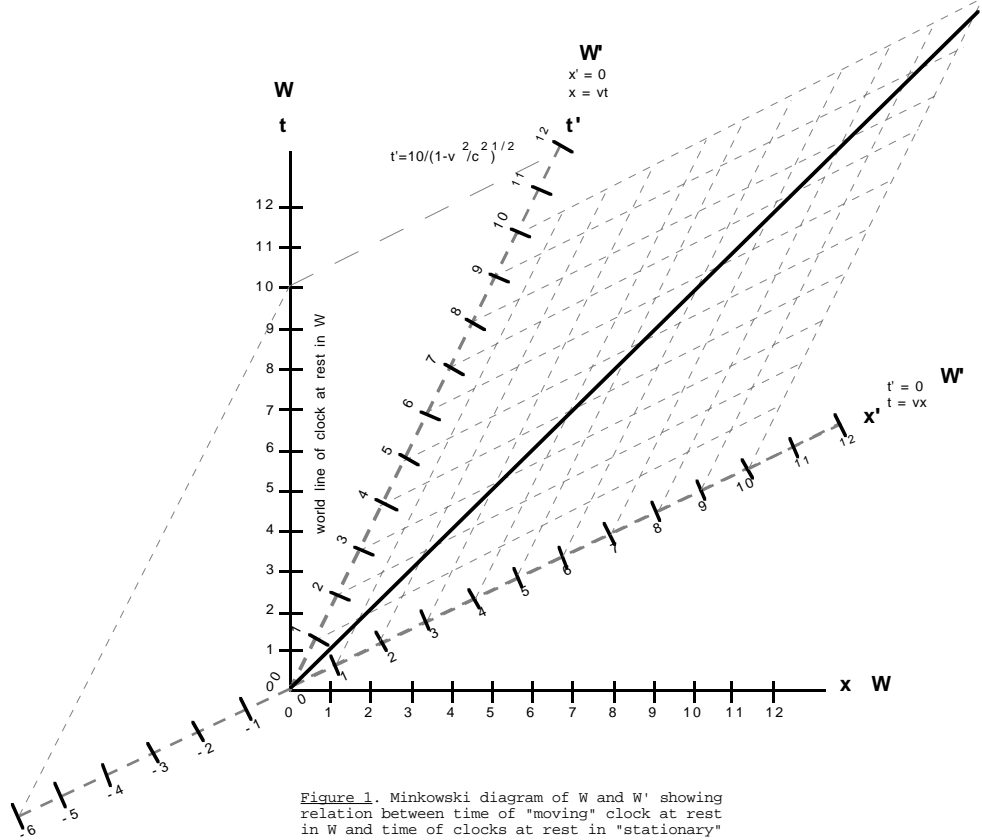

Figure 1. Minkowski diagram of W and W' showing relation between time of "moving" clock at rest in W and time of clocks at rest in "stationary" reference frame W'.

Figure 1 Minkowski diagram of W and W' showing relation between time of "moving" clock at rest in W and time of clocks at rest in "stationary" reference frame W'.

# On the Arbitrary Choice

relationship of temporal durations measured by observers at rest in their respective inertial reference frames and who use clocks at rest in their respective frames is given by equation 1. In this scenario, W' is the "stationary" inertial reference frame and W is the "moving" inertial reference frame.

Let the axis t from t = 0 to t = 10 represent the position of a clock at rest in W from t = 0 to t = 10 at the spatial origin of W, x = 0. In W', the times corresponding to t = 0 and t = 10 will be measured by different clocks that are at rest in W'. The clock at rest in W' at x = 0, t = 0 in W will be at x' = 0 in W'. The clock at rest in W' at x = 0, t = 10 in W will be at x' = -6 in W'. (Remember that in W' the clock at rest in W is moving with the same uniform, translational velocity that W is moving relative to W'.) In order to determine the duration in W' corresponding to a duration of 10 units in W, one needs to extend a line parallel to the spatial axis in W', namely x', and intersecting x = 0 and t = 10 in W, to see where it intersects with the time axis for W', t'. (This line represents the temporal coordinate, t', in W' that corresponds to the point x = 0 and t = 10 in W. The intersection of this line, parallel to the x' axis, with the axis t' yields the time coordinate in W' corresponding to t = 10 and x = 0 in W.) The t' axis is intersected at $t' = 10/(1 - v^2/c^2)^{1/2}$. As noted, the time and space coordinates of the origins of W and W' overlap, and thus t' = 0 and x' = 0 in W' corresponds to t = 0 and x = 0 in W. The difference between $t' = 10/(1 - v^2/c^2)^{1/2}$ and t' = 0, namely $\Delta t' = 10/(1 - v^2/c^2)^{1/2}$ is the amount of time in W' corresponding to $\Delta t$ = 10 units in W. In Figure 1, it is readily seen that the amount of time elapsed in W' corresponding to the elapse of 10 units in W is over 10 units.

Consider the *reverse* circumstance as portrayed in Figure 2 and which is a pictorial representation of equation 2. Here an observer in W considers the time measured by a clock at rest in W'. With regard to W, the clock at rest in W' moves with a uniform translational velocity v (the sign is changed because of the reversal in direction of the velocity). In this scenario, W is the "stationary" inertial reference frame and W' is the "moving" inertial reference frame. In this case, let the axis t' from t' = 0 to t' = 10 represent the position of a clock at rest in W' from t' = 0 to t' = 10 at the spatial origin of W', x' = 0. In W, the times corresponding to t' = 0 and t' = 10 will be measured by different clocks that are at rest in W. In order to determine the duration in W corresponding to a duration of 10 units in W', one needs to extend a line parallel to the spatial axis in W, namely x, and intersecting x' = 0 and t' = 10 in W', to see where it intersects with the time axis for W, t. (This line represents the temporal coordinate, t, in W that corresponds to the point x' = 0 and t' = 10 in W'. The intersection of this line parallel to x with the axis t yields the time coordinate in W corresponding to t' = 10 and x' = 0 in W'.) The t axis is intersected at $t = 10/(1 - v^2/c^2)^{1/2}$. As noted, the time and space coordinates of the origins of W and W' overlap, and thus t = 0 in W corresponds to t' = 0 and x' = 0 in W'. The difference between $t = 10/(1 - v^2/c^2)^{1/2}$ and t = 0, namely $\Delta t = 10/(1 - v^2/c^2)^{1/2}$ is the amount of time in W corresponding to $\Delta t'$ = 10 units in W'. In Figure 2, it is readily seen that the amount of time elapsed in W corresponding to the elapse of 10 units in W' is over 10 units.

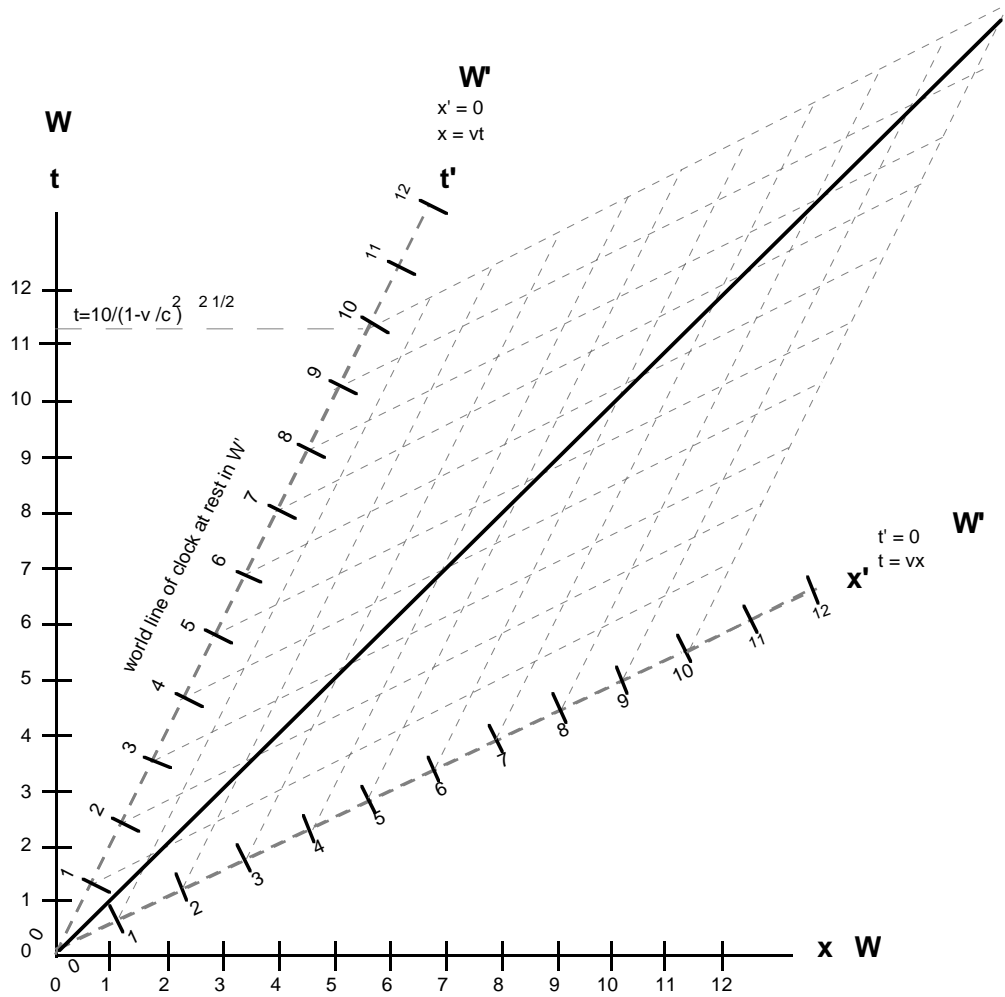

Figure 2. Minkowski diagram of W and W' showing relation between time of "moving" clock at rest in W' and time of clocks at rest in "stationary" reference frame W.

The two circumstances discussed above, in essence, reflect the conundrum concerning the relation between temporal durations in inertial reference frames in uniform translational motion relative to one another. At first, it seems there is no problem at all. In each scenario, a clock is designated at rest in one of the inertial reference frames (e.g., Q) and is considered moving with uniform translational velocity relative to an observer at rest in the other inertial reference frame (e.g., P). This all seems very straightforward. One is dealing with a clearly designated clock, the particular motion of which is clearly delineated for observers at rest in their respective inertial reference frames. The problem comes when one notes that the time of this clock is measured by the observer in P for whom the clock is moving in a uniform translational manner by using at least two clocks that are at rest in this observer's inertial reference frame. The problem arises because an observer at rest in the inertial reference frame Q which has what the observer in P considers the "moving" clock instead considers his own reference frame Q in which he is at rest the "stationary" reference frame and the other inertial reference frame P, including the clocks at rest in it, moving in a uniform translational manner relative to his reference frame. In terms of the modification of the twin paradox at issue in this paper, both twins correctly maintain in terms of the special theory that the other twin is aging more slowly.

*4.    The Relations Between Spatial Lengths in Inertial Reference Frames*

A similar effect occurs for spatial length as well. The relation for spatial lengths in the two inertial frames W and W' for the circumstance where the observer at rest in W' considers the length of a rod at rest in W is:

$\Delta x' = \Delta x(1 - v^2/c^2)^{1/2}$  (3).

# On the Arbitrary Choice

In this scenario, W' is the "stationary" inertial reference frame and W is the "moving" inertial reference frame. For the circumstance where the observer at rest in W considers the length of a rod at rest in W' and aligned along the direction of relative motion of W and W', the relation is:

$$\Delta x = \Delta x'(1 - v^2/c^2)^{1/2} \quad (4).$$

In this scenario, W is the "stationary" inertial reference frame and W' is the "moving" inertial reference frame.

As with the case of time and clocks, the analysis would seem very straightforward. A rod can be at rest in one inertial reference frame and moving with a uniform translational velocity relative to another inertial reference frame. If the rod is at rest in W, then the relation between the lengths of the rod in W and W' is given by equation 3. (As noted, in this scenario, W' is the "stationary" inertial reference frame and W is the "moving" inertial reference frame.) This circumstance is depicted in Figure 3. If a rod of length 10 units is at rest in W, the length of this rod is less than 9 units in W'. The dashed line represents part of the world line of the end of the rod that is situated at $x = 10$ when $t = 0$ in W where it is at rest. In order to measure the length of the rod in W', the "stationary" reference frame, the coordinates corresponding to the ends of the rod must be determined simultaneously in W'. Since it is known that the spatial and temporal origins of W and W' overlap, it is known that the end of the rod situated at $x = 0$ and $t = 0$ in W will have the coordinates $x' = 0$ and $t' = 0$ in W'. In W', the spatial coordinates of the other end of the rod corresponding to $t' = 0$ can be determined by the intersection of the world line of the rod with the x' axis in W'. This point, as shown in Figure 3, is between 8 and 9 units.[3]

If, instead, the rod is at rest in W', then the relation between the lengths of the rod in W and W' is given by equation 4. In this scenario, W is the "stationary" inertial reference frame and W' is the "moving" inertial reference frame. This circumstance is illustrated in Figure 4.[4]

The problem arises, though, that one can argue the relativity of simultaneity and thus derive the Lorentz coordinate transformation equations, upon which equations 3 and 4, as well as equations 1 and 2, are dependent, beginning with either of the inertial reference frames in uniform translational motion relative to one another. It is because one can begin the argument with either reference frame and establish simultaneity in either of the inertial reference frames first, which is deemed the "stationary" inertial reference frame,

On the Arbitrary Choice

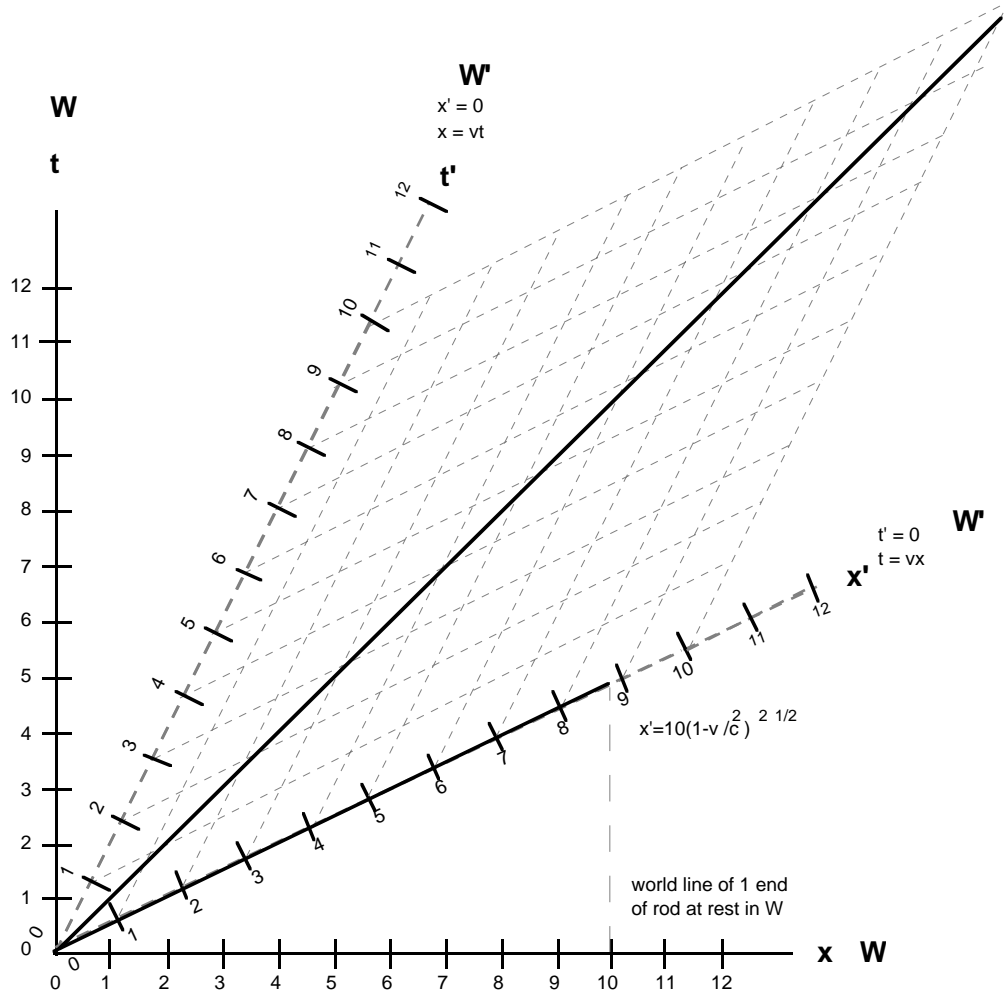

Figure 3. Minkowski diagram of W and W' showing relation between spatial length of rod at rest in W and spatial length of "moving" rod in "stationary" reference frame W'.

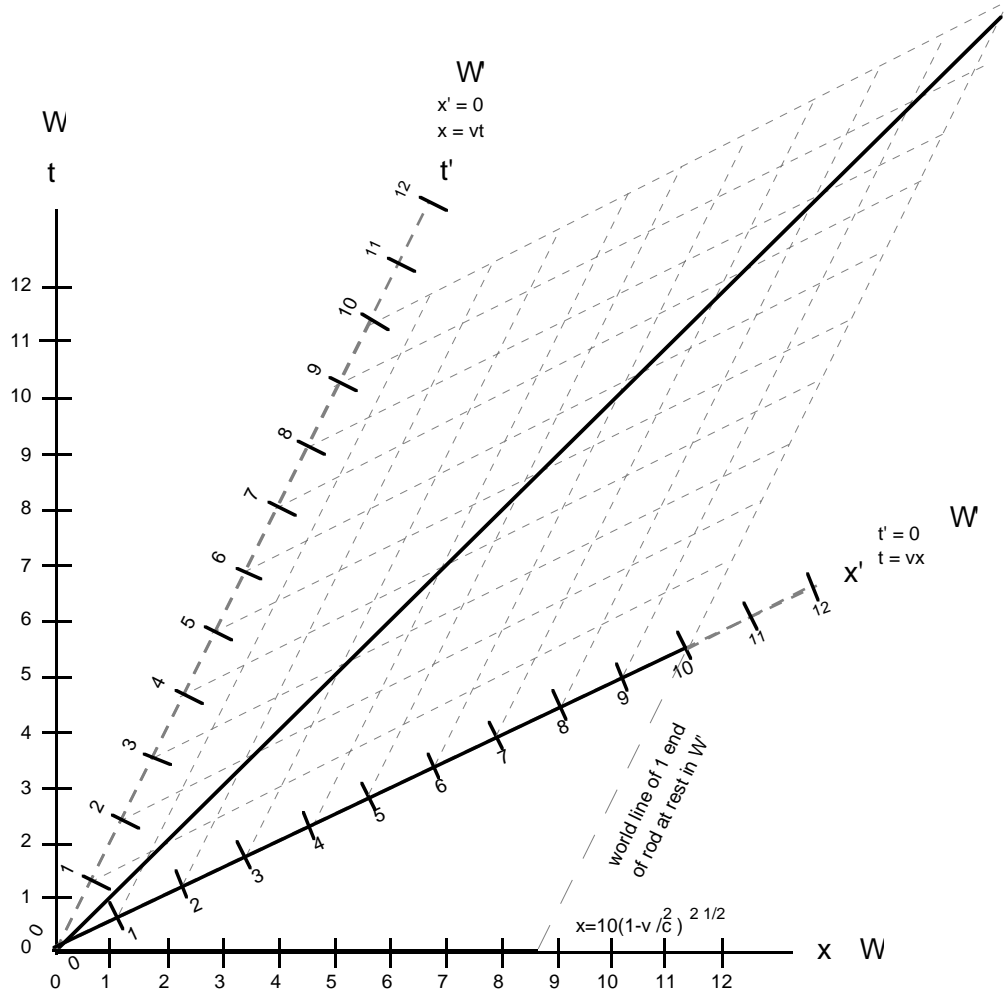

Figure 4. Minkowski diagram of W and W' showing relation between spatial length of rod at rest in W' and spatial length of "moving" rod in "stationary" reference frame W.

## On the Arbitrary Choice

that there are two reciprocal sets of equations relating spatial lengths and temporal durations for inertial reference frames in uniform translational motion relative to one another.

If it were not possible to begin the argument on the relativity of simultaneity and first establish simultaneity in either inertial reference frame, one of the inertial reference frames would be a preferred inertial reference frame, and the establishment of simultaneity, and thus time and as will be shown space, in this reference frame would be primary. This feature is associated with the fundamental tenet of the special theory that descriptions of physical phenomena from either of two inertial reference frames in uniform translational motion relative to one another are equivalent for observers at rest in their respective inertial reference frames. If this tenet did not hold, there would exist the possibility of unique descriptions of physical phenomena in the two inertial reference frames. The inertial reference frame in which simultaneity is first established would be the inertial reference frame from which the description of physical phenomena would be considered more fundamental.

A discussion of the underlying foundation of the relationships between spatial lengths and temporal durations in inertial reference frames in uniform translational motion relative to one another follows. First, how the relations for spatial lengths and temporal durations between inertial reference frames having a uniform translational velocity relative to one another are dependent on the Lorentz coordinate transformation equations is shown. Second, how the relations for spatial length and temporal duration and the Lorentz coordinate transformation equations depend on the relativity of simultaneity will be shown. Then, the arbitrary decision concerning which inertial reference frame is "stationary" and which "at rest" in the relativity of simultaneity will be demonstrated using Einstein's 1917 argument.[4] (As discussed previously, "stationary" refers to the reference frame where the argument concerning the relativity of simultaneity begins, where simultaneity, and thus time, is first established for one of the inertial reference frames in uniform translational motion relative to one another. "Moving" essentially refers to the reference frame where the time of the "stationary" reference frame is introduced in order to determine whether the criterion of simultaneity in the "moving" frame is met.)

It will be shown that the relativity of simultaneity can be argued with either inertial reference frame considered the "stationary" reference frame and the other reference frame considered the "moving" reference frame. This arbitrary characteristic of the argument on the relativity of simultaneity impacts the Lorentz coordinate transformation equations as they are dependent on the relativity of simultaneity in the special theory.  This arbitrary characteristic also impacts the relationships between the spatial lengths of a physical existent and the temporal durations of an occurrence in inertial reference frames moving in a uniform translational manner relative to one another and to other results in the special theory.

In the special theory, because the decision in which direction (i.e., which inertial reference frame is designated the "stationary" inertial reference frame as it is in this frame that simultaneity is first established) to argue the relativity of simultaneity is not constrained by any physical or mental factors, the decision is freely made by the individual making the argument.  The decision is a cognitive act on the part of this individual, and this cognitive act has consequences for events in the physical world.  This cognitive act and its direct relation to the physical world depends on, and also serves to explain, the feature of the physical world that an observer at rest in an inertial reference frame considers this frame "stationary".  This feature of the physical world allows for the fundamental tenet of the special theory that descriptions of physical phenomena from either of two inertial reference frames in uniform translational motion relative to one another are equivalent for observers at rest in their respective inertial reference frames.

*5.    Spatiotemporal Relations and the Lorentz Transformation Equations*

Two sets of reciprocal relations have been discussed that relate the temporal durations of an occurrence and the spatial lengths of a physical existent in inertial reference frames in uniform translational motion relative to one another. One set of relations (set 1) is comprised of equations 1 and 3 and is pictorially represented in Figures 1 and 3.  The other set of relations (set 2) is comprised of equations 2 and 4 and is pictorially represented in Figures 2 and 4.

Set 1 is most naturally derived from the Lorentz coordinate transformation equations 5 and 6.  The term most naturally means that the inertial reference frame designated the "stationary" frame and the inertial reference frame designated the "moving" frame in the derivation of the Lorentz coordinate transformation equations retain their roles in the spatial length and temporal duration relations in set 1.[5]  Equations 5 and 6 are:

$$x = (x' + vt')/(1 - v^2/c^2)^{1/2} \quad (5)$$

and

$$t = [t' + (v/c^2)x']/(1 - v^2/c^2)^{1/2} \quad (6).$$

Set 2 is most naturally derived from the Lorentz coordinate transformation equations:

$$x' = (x - vt)/(1 - v^2/c^2)^{1/2} \quad (7)$$

and

$$t' = [t - (v/c^2)x]/(1 - v^2/c^2)^{1/2} \quad (8).$$

The derivations of equations 2 and 4 using equations 7 and 8 follow immediately. The derivation of equation 3 using equations 7 and 8 is then carried out in the section, "How Spatiotemporal Relations Between Inertial Reference Frames Depend on the Relativity of Simultaneity." This derivation will show the significance of the relativity of simultaneity in the equations relating spatial extension of physical existents and temporal duration of occurrences in inertial reference frames in uniform translational motion relative to one another. It does so, though, by *switching* in midstream, so to speak, which inertial reference frame is the "stationary" reference frame and which the "moving" reference frame. It is shown that the Lorentz coordinate transformations derived with one inertial reference frame the "stationary" reference frame and the other inertial reference frame the "moving" inertial reference frame can be used to deduce the relation between spatial lengths and temporal durations for these reference frames where the "stationary" frame becomes the "moving" frame and the "moving" frame becomes the "stationary" frame. It should be emphasized that this is possible only because the relativity of simultaneity can be argued in *either* direction, that is with either inertial reference frame designated the "stationary" reference frame and the other inertial reference frame designated the "moving" inertial reference frame.

*5.1. The Derivation of Equation 2*

W and W' have been specified as one dimensional spatial coordinate systems moving in a uniform translational manner relative to one another along their respective spatial axes, x of W and x' of W'. This relative motion can be determined from *either* W or W' by observers O (at rest in W) and O' (at rest in W'). Allow that O considers W' to be in uniform, translational motion with the velocity v along the x and x' axes in the direction of increasing values of x and x'. In this scenario, W is the "stationary" inertial reference frame and W' is the "moving" inertial reference frame. Also, for O, at rest in W, W is the "stationary" inertial reference frame and W' is the "moving" inertial reference frame. Then the Lorentz transformation equations of special relativity for transforming space and time coordinates for W (represented by the variables x and t, respectively) and space and time coordinates for W' (represented by x' and t', respectively) are equations 7 and 8. (Equations 7 and 8 are the relevant equations instead of equations 5 and 6 because in the argument on the relativity of simultaneity, W is considered "stationary" by O and O considered W' to be "moving.") (These same circumstances will hold for the derivation of equation 4 as well.)

Multiplying both sides of equation 8 by $(1 - v^2/c^2)^{1/2}$ and adding $(v/c^2)x$ to both sides of this equation, one obtains: $t = t'(1 - v^2/c^2)^{1/2} + (v/c^2)x$. The goal is to determine $t_2 - t_1$ (i.e., the temporal duration in W) corresponding to the temporal duration measured by a "moving" clock, which is at rest in W' (i.e., at a particular location $x_1'$). For $t_1$ and $t_2$ in W, the last equation becomes:

$$t_1 = t_1'[1 - v^2/c^2]^{1/2} + (v/c^2)x_1$$

and

$$t_2 = t_2'[1 - v^2/c^2]^{1/2} + (v/c^2)x_2$$

where $x_1$ and $x_2$ are coordinates in W such that $x_1'$ is the particular value in W' corresponding to the values $x_1$ and $x_2$, and $t_1'$ and $t_2'$ are the corresponding values in W' to $t_1$ and $t_2$. Equation 7 can be expressed as $x = x'(1 - v^2/c^2)^{1/2} + vt$. Substituting for $x_1$ and $x_2$ in the above equations for $t_1$ and $t_2$, one obtains:

$$t_1 = [t_1'(1 - v^2/c^2)^{1/2}] + [[v/c^2][x_1'(1 - v^2/c^2)^{1/2} + vt_1]]$$

and

$$t_2 = [t_2'(1 - v^2/c^2)^{1/2}] + [[v/c^2][x_1'(1 - v^2/c^2)^{1/2} + vt_2]].$$

Now,

$$t_2 - t_1 = [[t_2'(1 - v^2/c^2)^{1/2}] + [[v/c^2][x_1'(1 - v^2/c^2)^{1/2} + vt_2]]] - [[t_1'(1 - v^2/c^2)^{1/2}] + [[v/c^2][x_1'(1 - v^2/c^2)^{1/2} + vt_1]]]$$

$$t_2 - t_1 = [[t_2' - t_1'][(1 - v^2/c^2)^{1/2}]] + [(v/c^2)(vt_2 - vt_1)]$$

$$t_2 - t_1 = [[t_2' - t_1'][(1 - v^2/c^2)^{1/2}]] + [(v^2/c^2)(t_2 - t_1)] \ .$$

Subtracting $[(v^2/c^2)(t_2 - t_1)]$ from both sides of the above equation, one obtains:

$$(t_2 - t_1) - [(v^2/c^2)(t_2 - t_1)] = [t_2' - t_1'][(1 - v^2/c^2)^{1/2}]$$

$$(t_2 - t_1)(1 - v^2/c^2) = [t_2' - t_1'][(1 - v^2/c^2)^{1/2}]$$

$$t_2 - t_1 = [[t_2' - t_1'][(1 - v^2/c^2)^{1/2}]]/(1 - v^2/c^2) \ .$$

Thus,

$$t_2 - t_1 = (t_2' - t_1')/(1 - v^2/c^2)^{1/2} \ .$$

With $\Delta t = t_2 - t_1$ and $\Delta t' = t_2' - t_1'$,

$$\Delta t = \Delta t'/(1 - v^2/c^2)^{1/2} \quad (2).$$

It should be kept in mind that the inertial reference frame which has been designated "stationary," W, could just as easily have been designated the "moving" inertial reference frame. W' could just as easily have been designated the "stationary" inertial reference frame instead of the "moving" reference frame. In this alternative, scenario, equation 1 would have been derived using equations 5 and 6.

*5.2  The Derivation of Equation 4*

Using the same circumstances involving W and W' specified for the derivation of equation 2, allow that the rod, aligned along the line of relative motion, in our example is at rest in W' and is thus moving with uniform, translational velocity v relative to O. As in the previous example, W is considered the "stationary" frame and W' is considered the "moving" frame. A rod at rest in W' is measured to be r units by O', who is at rest in W'. The length of this rod when measured by O is calculated by finding the absolute value of the difference of the rod's coordinates along the x axis (i.e., $x_1$ and $x_2$) when measurements of these coordinates are made simultaneously at the time $t_1$, with simultaneity defined specifically for W. The rod's length is $x_2 - x_1|$, where the enclosure |   | represents the absolute value of the enclosed difference. Since the length of the rod in W' is known, equation 7 can be used to determine the length of the rod in W. Multiplying both sides of equation 7 by $(1 - v^2/c^2)^{1/2}$ and adding vt to both sides of the equation, one obtains $x = x'(1 - v^2/c^2)^{1/2} + vt$. The goal is to determine $|x_2 - x_1|$ at a particular time $t_1$. (Since the coordinates of the end of the rod in W will be known at the same time $t_1$ in W, $|x_2 - x_1|$ is the length of the "moving" rod in W). For the particular coordinates $x_1$ and $x_2$ at a particular time $t_1$, the last equation becomes $x_1 = x_1'(1 - v^2/c^2)^{1/2} + vt_1$ and $x_2 = x_2'(1 - v^2/c^2)^{1/2} + vt_1$, where $x_1'$ and $x_2'$ are the corresponding $x'$ coordinates in W'. Now,

$$|x_2 - x_1| = |(x_2'(1 - v^2/c^2)^{1/2} + vt_1) - (x_1'(1 - v^2/c^2)^{1/2} + vt_1)|$$

$$|x_2 - x_1| = |x_2'(1 - v^2/c^2)^{1/2} - x_1'(1 - v^2/c^2)^{1/2} + vt_1 - vt_1|$$

$$|x_2 - x_1| = |(x_2' - x_1')(1 - v^2/c^2)^{1/2}|,$$

or with $\Delta x = |x_2 - x_1|$ and $\Delta x' = |x_2' - x_1'|$,

$$\Delta x = \Delta x'(1 - v^2/c^2)^{1/2} \quad (4).$$

As $|x_2' - x_1'| = r$ and is positive,

$$\Delta x = r(1 - v^2/c^2)^{1/2}.$$

Just as equation 1 of set 1 could have been derived from equations 5 and 6, instead of equation 2 of set 2 from equations 7 and 8, so equation 3 of set 1 could just as well have been derived from equation 5, instead of equation 4 of set 2 from equation 7. A summary of the circumstances for deriving set 1 and equations 5 and 6 is presented in Table 1 in Appendix 1. A summary of the circumstances for deriving set 2 and equations 7 and 8 is presented in Table 2 in Appendix 1.

These tables illustrate the point that the *same* concrete circumstances in the physical world can support both scenarios concerned with the relation between the spatial lengths of physical existents and the temporal duration of occurrences in inertial reference frames in uniform translational motion relative to one another. That is, there may be similarly constructed measuring rods and clocks at rest in each of these respective inertial reference frames. The question arises, and will be addressed later, as to the basis for the different empirical results regarding the spatial lengths of physical existents and the temporal durations of occurrences in these inertial reference frames if the same concrete circumstances in the physical world can support either scenario. Also, the lengths of the measuring rods at rest or the clocks at rest in an inertial reference frame will be found to depend on whether they are at rest in the "stationary" reference frame or the "moving" reference frame. This last feature of the special theory serves to emphasize that the spatiotemporal structure of the physical world does not depend on concrete circumstances in the physical world.

*6.    How Spatiotemporal Relations Between Inertial Reference Frames Depend on the Relativity of Simultaneity*

The central importance of the relativity of simultaneity to the Lorentz transformation for time coordinates, and thus to the functioning of clocks at rest in their respective inertial frames in uniform translational motion relative to one another, is shown by discussing the derivation of the length of a rod that is at rest in W and that is thus a moving rod for O' who is at rest in W'. This derivation will, of course, demonstrate the importance of the relativity of simultaneity to the functioning of measuring rods at rest in their respective inertial reference frames and also to the Lorentz transformation equation for space coordinates along the axis of uniform translational velocity of the inertial reference frames relative to one another.

As noted, W and W' have been specified as one dimensional spatial coordinate systems moving in a uniform translational manner relative to one another along their respective spatial axes. This relative motion can be determined from *either* W or W' by observers O (at rest in W) and O' (at rest in W'). Allow that O considers W' to be in uniform translational motion with the velocity v along the x and x' axes in the direction of increasing values of x and x'. That is, W is the "stationary" inertial reference frame and W' is the "moving" inertial reference frame in this scenario. Then the Lorentz transformation equations of special relativity for transforming space and time coordinates for W (represented by the variables x and t) and space and time coordinates for W' (represented by x' and t') are:

$$x' = (x - vt)/(1 - v^2/c^2)^{1/2} \quad (7)$$

and

$$t' = [t - (v/c^2)x]/(1 - v^2/c^2)^{1/2} \quad (8)$$

where v is the relative uniform translational motion of W and W' and c is the invariant velocity of light in all inertial frames of reference.[5]

Let the length of a measuring rod that is at rest in either W or W', and which is aligned along the line of relative motion of W and W', be measured by observers at rest with regard to the rod to be r units. In the special theory, the length of a similarly constructed and similarly aligned measuring rod at rest in the other inertial frame will also be r units when this length is determined by observers at rest in the other frame. But the length of the rod at rest in W', for example, will not be r units when determined by observers at rest in W for which the rod is moving at the uniform translational velocity v in the same direction as W' is moving relative to W. Rather, the length of the rod, aligned along the direction of relative motion of W and W' will be $r(1 - v^2/c^2)^{1/2}$ units for observers at rest in W (here the "stationary" frame), where v is the uniform translational velocity of one frame relative to the other, c is the constant and finite velocity of light in all inertial frames, and v is less than c.

*6.1    Simultaneity and Spatial Measurement*

Allow now that the measuring rod of concern is at rest in W, instead of W', and is thus moving with uniform translational velocity v relative to O'. As noted, this rod is measured to be r units by O. The length of this rod when measured by O' is calculated by finding the absolute value of the difference of the rod's coordinates along the x' axis (i.e., $x'_1$ and $x'_2$) when measurements of these coordinates are made simultaneously at the time $t'_1$, with simultaneity defined specifically for W'. (Here W' is the "stationary" frame and W is the "moving" frame.) The rod's length is $|x'_2 - x'_1|$, where the enclosure | | represents the absolute value of the enclosed difference. Since the length of the rod in W is known, the Lorentz transformation equations 7 and 8 can be used to determine the length of the rod in W'. For $\Delta x' = |x'_2 - x'_1|$, $\Delta x = r = |x_2 - x_1|$, and $\Delta t = t_2 - t_1$,

$$\Delta x' = (\Delta x - v\Delta t)/(1 - v^2/c^2)^{1/2} \quad (9)$$

$$\Delta t' = [\Delta t - (v/c^2)\Delta x]/(1 - v^2/c^2)^{1/2} \quad (10)$$

In equation 10, it can be seen that there has to be a difference $\Delta t = v/c^2(r)$ in order for $\Delta t' = 0$. With $\Delta t' = 0$ in equation 10, $\Delta x'$ can be derived using equation 9 because it is assured that the two ends of the rod are being measured simultaneously in W'. Since it is necessary that $\Delta t = (v/c^2)r$ in order that $\Delta t' = 0$, $\Delta t$ has a non-zero value whenever the inertial frames are moving relative to one another.

To find $\Delta x'$, given that $\Delta t = v/c^2(\Delta x)$:

$$\Delta x' = (\Delta x - v(v/c^2(\Delta x)))/(1 - v^2/c^2)^{1/2}$$

On the Arbitrary Choice

$$\Delta x' = (\Delta x - (v^2/c^2(\Delta x))/(1 - v^2/c^2)^{1/2}$$
$$\Delta x' = [(\Delta x)(1 - v^2/c^2)]/(1 - v^2/c^2)^{1/2}$$
$$\Delta x' = [(\Delta x)(1 - v^2/c^2)]/(1 - v^2/c^2)^{1/2}$$
$$\Delta x' = (\Delta x)(1 - v^2/c^2)^{1/2} \quad (3).$$

Equation 10 is an expression of the relation between temporal durations of an occurrence in these inertial reference frames, and the term $v/c^2(r)$ reflects the relativity of simultaneity in the special theory. Further, equation 8 of the Lorentz coordinate transformation equations, from which equation 10 is simply derived, contains the term $v/c^2(x)$. It is also shown that the relativity of simultaneity is central to the relation between spatial extension in inertial reference frames in uniform translational motion relative to one another. It is clear from an inspection of equations 5 through 8 that space and time are dependent on one another in the Lorentz coordinate transformation equations. Not surprisingly, the Lorentz coordinate transformation equations themselves depend on the relativity of simultaneity.[5,6]

The relativity of simultaneity is built into the term $(1 - v^2/c^2)^{1/2}$. If c were arbitrarily great instead of invariant and finite, equations 6 and 8 would be t = t', the Galilean coordinate transformation for time, which is based on absolute simultaneity for inertial reference frames in uniform translational motion relative to one another.

Equations 7 and 8 of the Lorentz coordinate transformation equations are derived with W the "stationary" frame and W' the "moving" frame. What has been done in the derivation of equation 3 here is to <u>switch</u> which inertial reference frame is "moving" and which "stationary" by making $\Delta t' = 0$ through setting $\Delta t = v/c^2(r)$ instead of $\Delta t = 0$ as was done in the derivation of $\Delta x = \Delta x'(1 - v^2/c^2)^{1/2}$ (4) using equation 7. In setting $\Delta t' = 0$, one is concerned with the length of the "moving" rod, at rest in W (the "moving" frame), as measured from W' (the "stationary" frame). In setting $\Delta t = 0$ in the derivation of equation 4 from equation 7, one is concerned with the length of the "moving" rod, at rest in W' (the "stationary" frame), as measured in W (the "moving" frame). It should be noted that this analysis in which equation 3 was derived using equations 7 and 8 could instead have been argued using equations 5 and 6. Then, equation 4 would have been the result.

In the foregoing analysis, it has been shown that spatiotemporal relations between inertial reference frames in uniform translational motion

# On the Arbitrary Choice

relative to one another depend on which frame is considered the "stationary" inertial reference frame and which the "moving" inertial reference frame. Moreover, the importance of the relativity of simultaneity to these spatiotemporal relations and the Lorentz coordinate transformation equations has been shown.

Now Einstein's argument on the relativity of simultaneity needs to be presented, and it needs to be shown that the argument can be made from either direction, with either inertial reference frame the "stationary" frame while the other frame is designated the "moving" inertial reference frame once the "stationary" frame in a scenario has been designated. Furthermore, the arbitrary choice in which direction the relativity of simultaneity is argued needs to be shown. The point needs to be elaborated that these directions, *each carrying their own distinct empirical results* (as has been shown concerning the relationship between temporal durations of occurrences and spatial lengths of physical existents for inertial reference frames in uniform translational motion relative to one another), are reflected in observers being at rest in their respective inertial reference frames and experiencing their inertial reference frame as "stationary" and concluding that the other inertial reference frame is "moving." It is this experience on the part of observers that allows there to be an arbitrary decision on the part of the individual considering the argument on the relativity of simultaneity or, for that matter, the relationship between spatial lengths of objects and temporal durations of occurrences in inertial reference frames in uniform translational velocity relative to one another.

Just as there can be no physical constraint pointing to one or the other of the inertial reference frames being favored in terms of the description of physical phenomena, so there can be no mental constraint as well. That observers at rest in their inertial reference frames consider their respective frames the "stationary" reference frame indicates there is no mental constraint. The circumstances, both physical and mental, are presented pictorially in Tables 1 and 2, and it can be seen that the same concrete circumstances in the physical world as well as the mental circumstances of the observers at rest in their inertial reference frames are the same in both tables.

There are those, such as Grünbaum[6] and Winnie[7,8], who maintain essentially that the relativity of simultaneity is not necessary to derive other results of the special theory. As Franzel[9] noted, those adhering to a conventionalist position, such as Grünbaum, maintain that the round trip velocity of light used in establishing simultaneity in the special theory does not necessarily entail that the velocity of this light is in each direction the same. That is, even though the averaged, round trip velocity agrees with the empirically determined velocity of light in vacuo, its one-way velocity need not be in agreement with the empirically determined value. Franzel also noted that a kind of "practical" absolute simultaneity is possible as an alternative to the relativity of simultaneity developed by Einstein. The point missed by such adherents to these other positions is the theoretical simplicity and yet fundamental and far reaching ramifications of the argument on the relativity of simultaneity as regards the special theory. It is not difficult to show that it is the relativity of simultaneity, as opposed to the absolute simultaneity of the kinematics underlying Newtonian mechanics, that distinguishes the results of the special theory from the results of Newtonian mechanics. It has been shown, for example, in the present paper how the relativity of simultaneity is central to the Lorentz coordinate transformation equations and the spatiotemporal relations that underlie the other results of the special theory. It should be emphasized that the concern in the present paper is not with attempting to refute the significance of the relativity of simultaneity in the special theory. The concern is with investigating certain implications of the relativity of simultaneity in the special theory.

7. *Einstein's 1917 Argument on the Relativity of Simultaneity*

The relativity of simultaneity will be argued using a slightly modified version of Einstein's 1917 gedankenexperiment (i.e., thought experiment).[10] The gedankenexperiment involves a railway train that is moving with a uniform translational velocity v along an embankment. Both the train and embankment are considered inertial frames of reference.

An observer ($O_t$) is located on the railway train midway between the ends of the train. In addition, an observer ($O_e$) is located midway between the points on the embankment corresponding to the ends of the train just when two lightning flashes strike the ends of the train. Let the motion of the train and the light flash in event A have the same direction in the frame of the embankment, and the motion of the train and the light flash in event B have opposite directions in the frame of the embankment.

Einstein wrote concerning the special theory that two events, which each give off a flash of light, may be considered simultaneous in an inertial frame of reference when an observer located midway between the spatial locations of the events observes these flashes of light emitted in both events at

# On the Arbitrary Choice

the same time.(11)  According to Einstein's argument, due to the postulated and empirically validated invariant velocity of light, the two lightning flashes in events A and B meet at the observer O who is located at the midpoint of the embankment.  The events A and B in which these flashes occurred are considered simultaneous in the frame of the embankment, in accordance with Einstein's definition.  (Please see Figure 5.)  Here, the embankment is the "stationary" inertial reference frame, the reference frame where the argument on the relativity of simultaneity begins.  It is the inertial reference frame in which simultaneity, and thus time, is first established for one of the inertial reference frames.  Einstein described this scenario in the following way:

> When we say that the lightning strokes A and B are simultaneous with respect to the embankment, we mean: the rays of light emitted at the places A and B, where the lightning occurs, meet each other at the mid-point M of the length A --> B of the embankment.  But the events A and B also correspond to positions A and B on the train. Let M' be the mid-point of the distance A --> B on the travelling train.  Just when the flashes[1] of lightning occur [that result in the flashes of light noted above], this point M' [the midpoint of the train] naturally coincides with the point M [the midpoint of that part of the embankment corresponding to the "moving" train], but it moves...with the velocity $v$ of the train.(12)

To which reference frame are the flashes of lightning referred to first?  In which reference frame are they used in an attempt to establish simultaneity, and thus also time, first in accordance with Einstein's definition?  Einstein answered the question in writing that "the rays of light at the places A and B...meet each other at the mid-point M of the length A --> B of the embankment."  He also answered these questions when he wrote in the footnote appended to the phrase, "Just when the flashes[1] of lightning occur":

> [1] As judged from the embankment.(13)

It is the time in the reference frame of the embankment which is given priority and established first in Einstein's argument on the relativity of simultaneity.

But, Einstein argued, the situation for the observer $O_t$ on the train is different. Einstein wrote:

# On the Arbitrary Choice

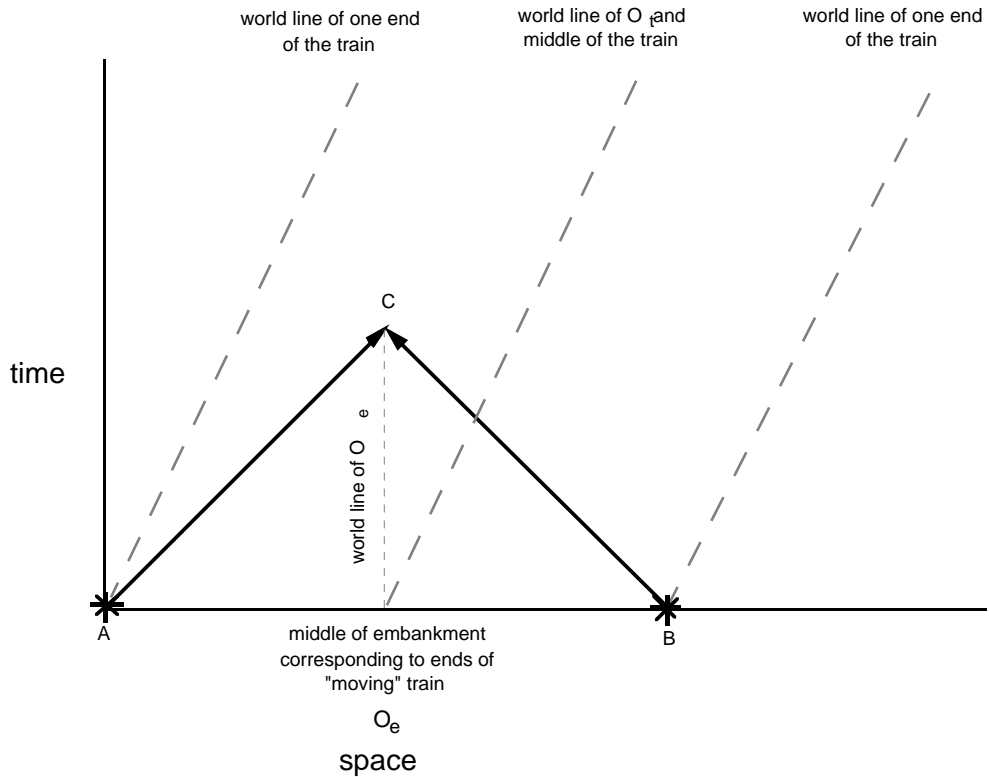

Time and space axes for the inertial reference frame of the embankment.

✹ are the events at which light flashes are emitted from the postions of the embankment corresponding to the ends of the train with the frame of the embankment considered the "stationary" reference frame.

Figure 5. Simultaneity in the inertial reference frame of the embankment according to Einstein's train gedankenexperiment.

# On the Arbitrary Choice

> Now in reality (considered with reference to the railway embankment) he [the observer on the train] is *hastening* [italics added] towards the beam of light coming from B, whilst he is *riding on ahead* [italics added] of the beam of light coming from A. Hence the observer will see the beam of light emitted from B earlier than he will see that emitted from A. Observers who take the railway train as their reference-body must therefore come to the conclusion that the lightning flash B took place earlier than the lightning flash A.(14)

Where the frame of the embankment is the "stationary" frame, for $O_t$, the lightning flashes have different effective velocities, depending on whether the particular beam is moving in the same or opposite direction to that of the train. (The measured velocity of light by $O_t$ is, of course, the same finite and invariant value it has in all inertial reference frames. It is this finite and invariant value that is central to Einstein's criterion for simultaneity in an inertial reference frame.) In the case where the train has the uniform velocity v and the lightning flashes have the invariant velocity c for the observer O on the embankment, the light flash from A has the effective velocity c - v and the light flash from B has the effective velocity c + v relative to the observer on the train.[7] In Figure 6, it can be seen that the flashes of light do not meet at the observer $O_t$ who is at rest midway between the ends of the train. When $O_t$ applies Einstein's criterion for simultaneity for an inertial reference frame to these light flashes, he finds the criterion is not met. Thus, Einstein concluded that two occurrences which are simultaneous for the observer at rest in the inertial frame of the embankment are not simultaneous for the observer at rest in the inertial frame of the train. (In that the observer on the train "is hastening towards the beam from light from B, whilst he is riding on ahead of the beam of light coming from A," Einstein again confirmed that the observer on the train is using the time of the reference frame of the embankment to determine whether simultaneity of the reference frame of the embankment holds in the reference frame of the train as well.)

In Einstein's scenario, the train is the "moving" inertial reference frame, the reference frame where the time of the "stationary" reference frame is applied through the use of the terms c - v and c + v to determine whether Einstein's criterion for simultaneity in the reference frame of the train is met. In terms of the space of the inertial reference frame of the railway embankment, the midpoint of the train will be displaced by v∆t, where v is the uniform

## On the Arbitrary Choice

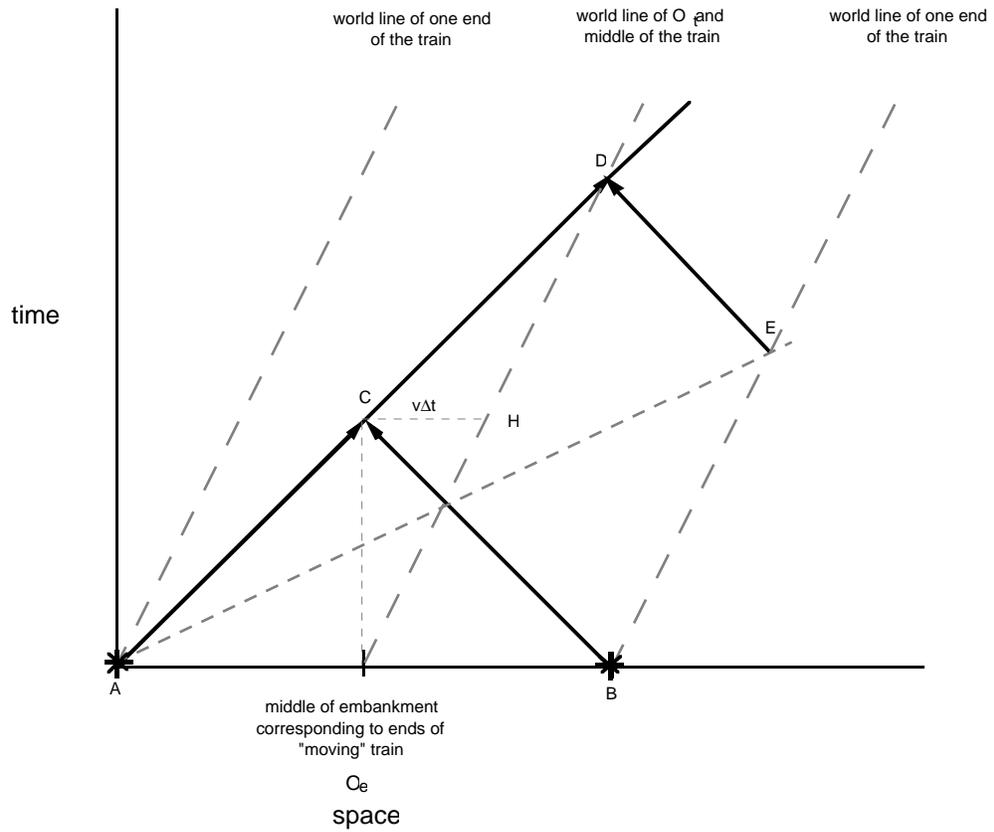

Time and space axes for the inertial reference frame of the embankment.

✱ are the events at which light flashes are emitted from the postions of the embankment corresponding to the ends of the train with the frame of the embankment considered the "stationary" reference frame.

Figure 6. Simultaneity in the inertial reference frames of the embankment and the train with the embankment considered the "stationary" reference frame and the train the "moving" reference frame.

# On the Arbitrary Choice

translational velocity of the train relative to the embankment and $\Delta t$ is the time in the reference frame of the embankment taken by the rays of light that strike the ends of the train to reach O at the midpoint of the embankment.

Further discussion is needed on this last point concerning the arbitrary nature of deciding which frame is "stationary" and which is "moving" for the purpose of arguing the relativity of simultaneity. The argument presented above concerning the relativity of simultaneity may be applied almost exactly to the situation in the train gedankenexperiment where the observer at rest on the train is considered at rest in the "stationary" inertial reference frame and the observer at rest on the embankment is considered at rest in the "moving" inertial reference frame. Indeed, the only two changes are: 1) the *switch of roles* as to which frame is the "stationary" frame and which the "moving" frame (i.e., in which inertial reference frame simultaneity is first established), and 2) the reversal in direction of the velocity of the embankment and the train relative to one another.[8] In the scenario where the train is considered the "stationary" frame and the embankment the "moving" frame, because of the postulated and empirically validated invariant velocity of light, the two light flashes meet at the observer $O_t$ located midway on the train. The motion of the embankment and the light flash from B have the same direction in the frame of the train, and the motion of the embankment and the light flash from A have opposite directions in the frame of the train. (Please see Figure 7.)

The lightning flashes are considered simultaneous in the frame of the train (in this scenario, the "stationary" reference frame), in accordance with Einstein's definition of simultaneity. As deduced by the "stationary" observer on the train, the light flash in A has the velocity $c + v$ relative to the "moving" observer on the embankment and the light flash in B has the velocity $c - v$ relative to the observer on the embankment. In Figure 7, it can be seen that the flashes of light do not meet at the observer $O_e$ who is at rest midway between the ends of the embankment. When $O_e$ applies Einstein's criterion for simultaneity for an inertial reference frame to these light flashes, he finds the criterion is not met. Similar to the first scenario, it can be concluded that two occurrences which are simultaneous for the observer at rest in the inertial frame of the train are not simultaneous for the observer at rest in the inertial frame of the embankment.

There is nothing that points in any way to an individual arguing the relativity of simultaneity in one direction or the other, that is with one of the reference frames more likely than the other to be designated the "stationary"

# On the Arbitrary Choice

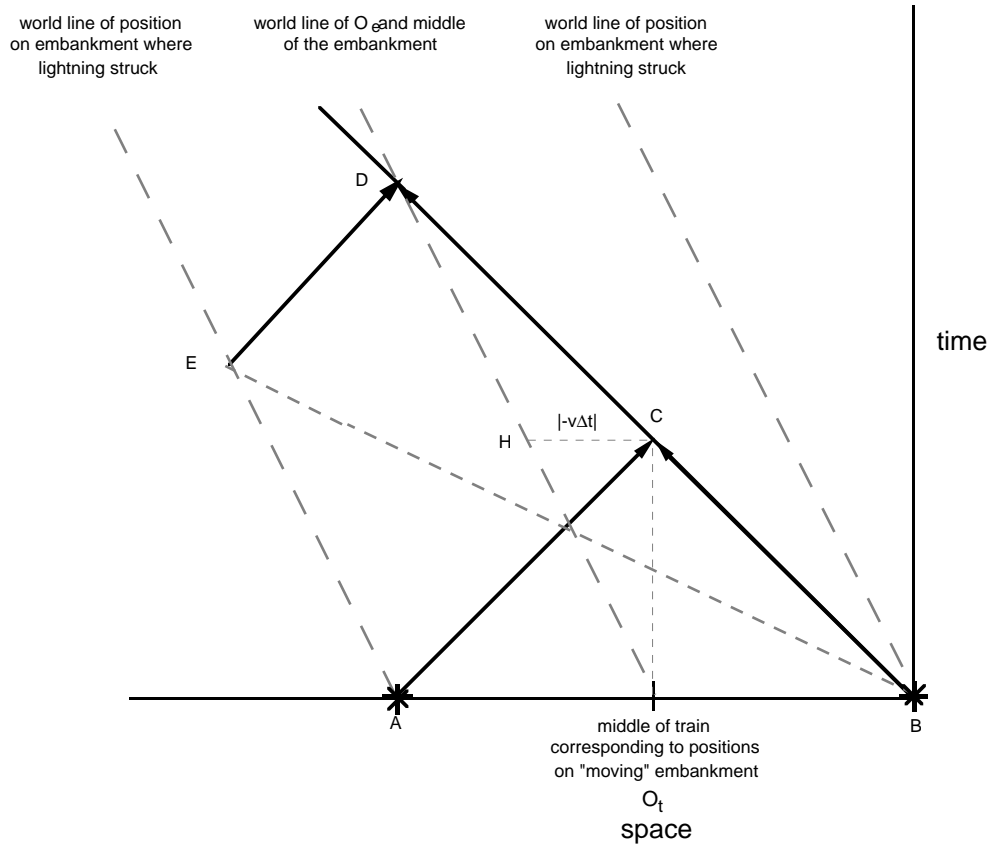

Time and space axes for the inertial reference frame of the train.

✱ are the events at which light flashes are emitted from the ends of the train corresponding to positions on the embankment with the frame of the train considered the "stationary" reference frame.

Figure 7. Simultaneity in the inertial reference frames of the embankment and the train with the embankment considered the "stationary" reference frame and the embankment the "moving" reference frame.

frame. If one were more likely, the fundamental tenet of the special theory that there is no preferred reference frame as concerns the description of physical phenomena for observers at rest in their respective inertial reference frames in uniform translational motion relative to one another would be violated. There would be a preferred inertial reference frame, and it would be the frame designated the "stationary" frame. There is thus a *free choice* on the part of the individual making the argument on the relativity of simultaneity as to which direction the argument should proceed. In that this choice is not limited by any physical factor in the special theory in any way, it can be said that the choice is fundamentally a cognitive one, not reducible to some physical substrate, and that this cognitive choice is reflected in distinct sets of empirical results found when certain measurements are taken in the physical world. Whichever direction is chosen, this direction will correspond to a distinct set of results obtained by an observer at rest in his inertial reference frame and which he considers the "stationary" reference frame. It should be noted that the assumption, supported empirically, that an observer at rest in his inertial reference frame considers this reference frame "stationary" and the other inertial reference frame "moving" supports the premise that there is no mental factor that prompts an individual to argue the relativity of simultaneity in one direction or the other.

*8.     The Arbitrary Decision in the Argument on the Relativity of
        Simultaneity*

The basic point is this. If the railway embankment is considered the "stationary" reference frame, then in Einstein's argument simultaneity is *first* established in this inertial reference frame *in terms of the motion of light*. The light flashes leaving from the points of the embankment corresponding to the ends of the train meet midway on the embankment between these two points. As concerns the train, these light flashes certainly do not meet midway on the train. This is seen in an inspection of Figure 5. As noted, simultaneity in the reference frame of the embankment is defined as the meeting at C of the light flashes represented by AC and BC. As can be seen in Figure 6, simultaneity in the train, the "moving" reference frame, is achieved when the flash leaving from the end of the train in event A meets at event D what appears to be a new flash leaving from the other end of the train in event E. E is an event on the x' axis. That the event D represents the light flashes AD and ED meeting midpoint on the train is evident from an inspection of Figure 8, which essentially is

# On the Arbitrary Choice

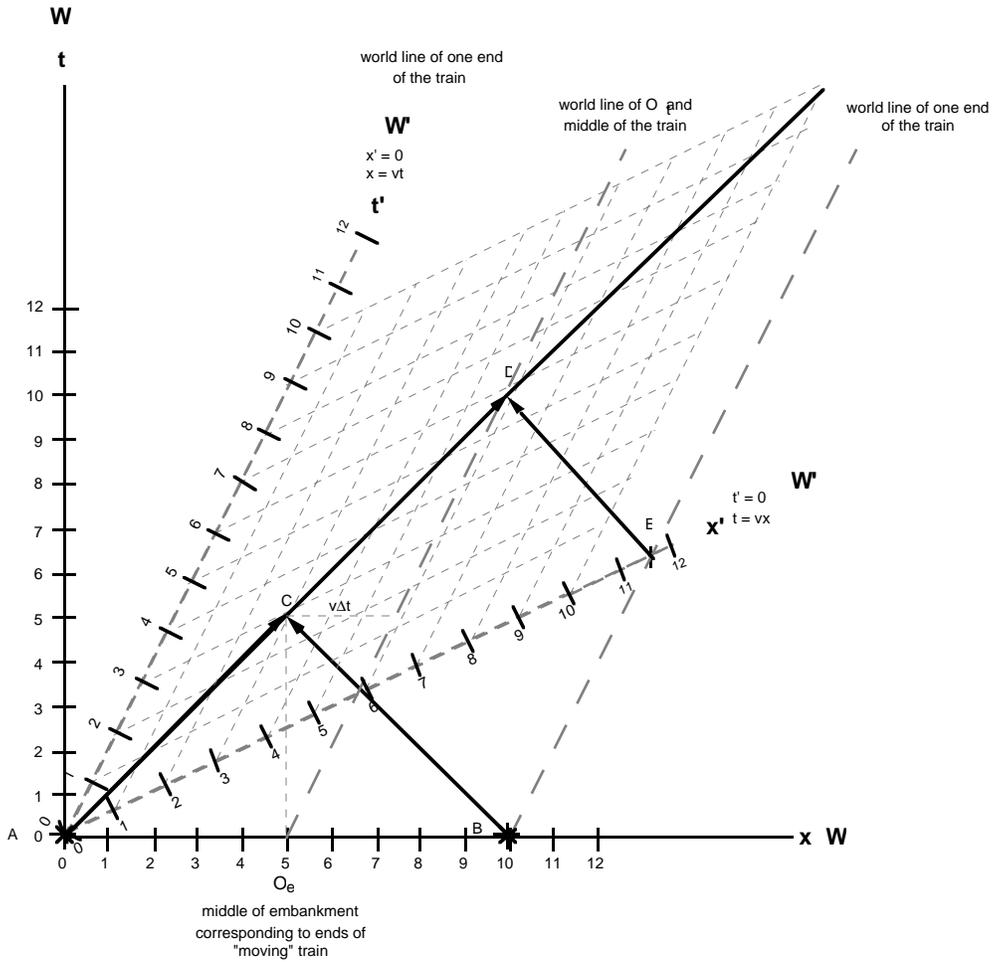

✳ are the events at which light flashes are emitted from the postions of the embankment corresponding to the ends of the train with the frame of the embankment considered the "stationary" reference frame.

Figure 8. Minkowski diagram of W and W' showing simultaneity in the inertial reference frames of the embankment (W) and the train (W') with the embankment considered the "stationary" frame.

# On the Arbitrary Choice

Figure 6 on which coordinate grids for the inertial reference frames of the embankment and the train have been added.

*8.1   The Spatial Length Along the Axis of Motion
of Physical Existents at Rest in the "Stationary"
and "Moving" Reference Frames*

One can see in Figure 8 that the length of the train measured by observers at rest on the train is not 10 units but more than 10 units, specifically $10/(1 - v^2/c^2)^{1/2}$ units, because the light flashes AD and DE meet Einstein's criterion that lights flashes from the ends of the train meet the observer at rest located midpoint on the train when the train is considered the "moving" inertial reference frame.  In greater detail, when the train is the "moving" reference frame, the events A and D correspond to the ends of the train at the same time in the reference frame of the train because the light flashes AD and DE meet Einstein's criterion for simultaneity.  Since A and D correspond to the ends of the train at the same time, the length from A to D is a measurement of the length of the train for observers at rest in the "moving" reference frame of the train.  This length, $10/(1 - v^2/c^2)^{1/2}$ units, is established in the argument on the relativity of simultaneity after length is established in the "stationary" frame of the embankment where light flashes occur in events A and B that are equidistant from the observer at rest in the "stationary" frame of the embankment.

Consider the reverse scenario, where the train is considered the "stationary" reference frame and the embankment is considered the "moving" reference frame.  The light flashes are the same physical phenomena as in Einstein's scenario.  In the reverse scenario, the light flashes AH and GH in Figure 9 are used to establish simultaneity in the inertial reference frame of the train first in the argument on the relativity of simultaneity.  The light flashes AD and DE, found in Figure 8, do not appear to be those flashes which are used to establish simultaneity in the reference frame of the train in Figure 9.  One would think that the same light flashes that were used to establish simultaneity in Einstein's scenario would be used in the reverse scenario being discussed.  But in comparing Figures 8 and 10 (where the space and time axes for the inertial reference frame of the embankment are drawn in), one can see even more clearly that this does not appear to be the case.

In the reverse scenario, the circumstances that had applied to the "moving" reference frame of the train now apply to the "moving" reference frame of the embankment and the circumstances that had applied to the

## On the Arbitrary Choice

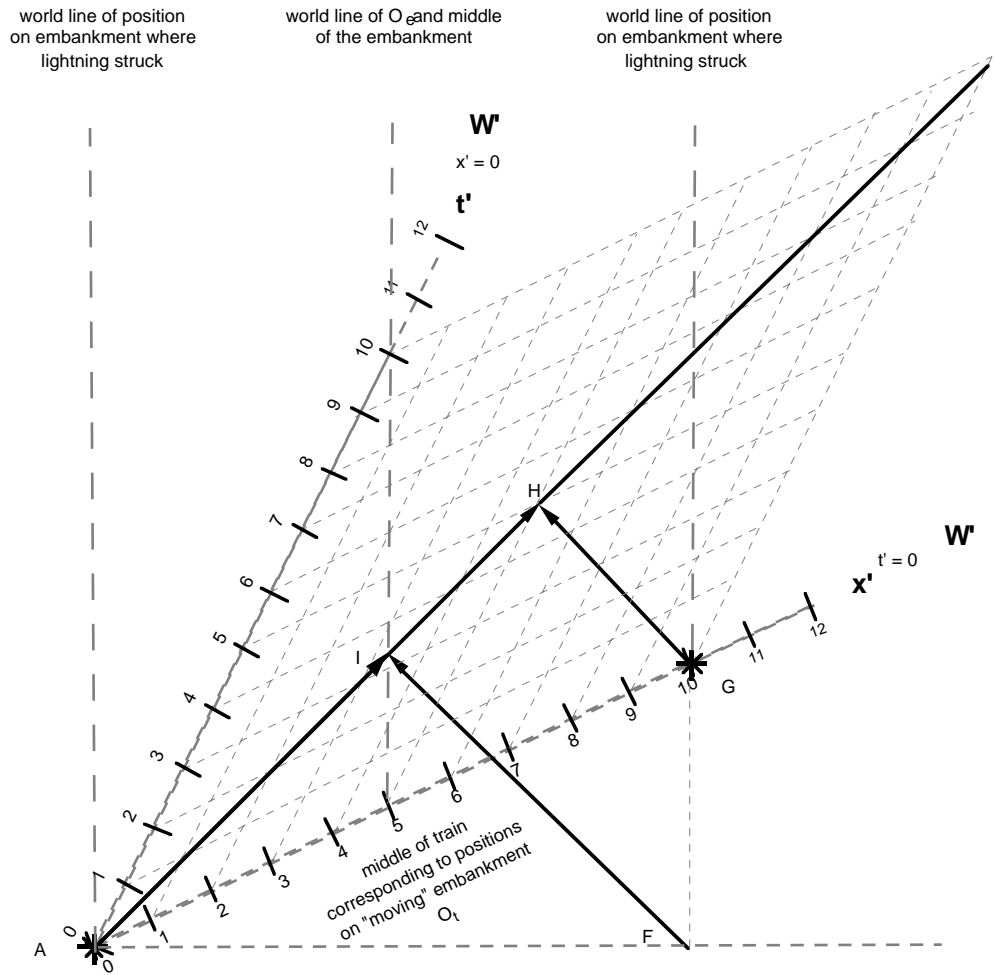

✱ are the events at which light flashes are emitted from the ends of the train corresponding to positions on the embankment with the frame of the train considered the "stationary" reference frame.

Figure 9. Simultaneity in the inertial reference frames of the train and the embankment with the train considered the "stationary" frame.

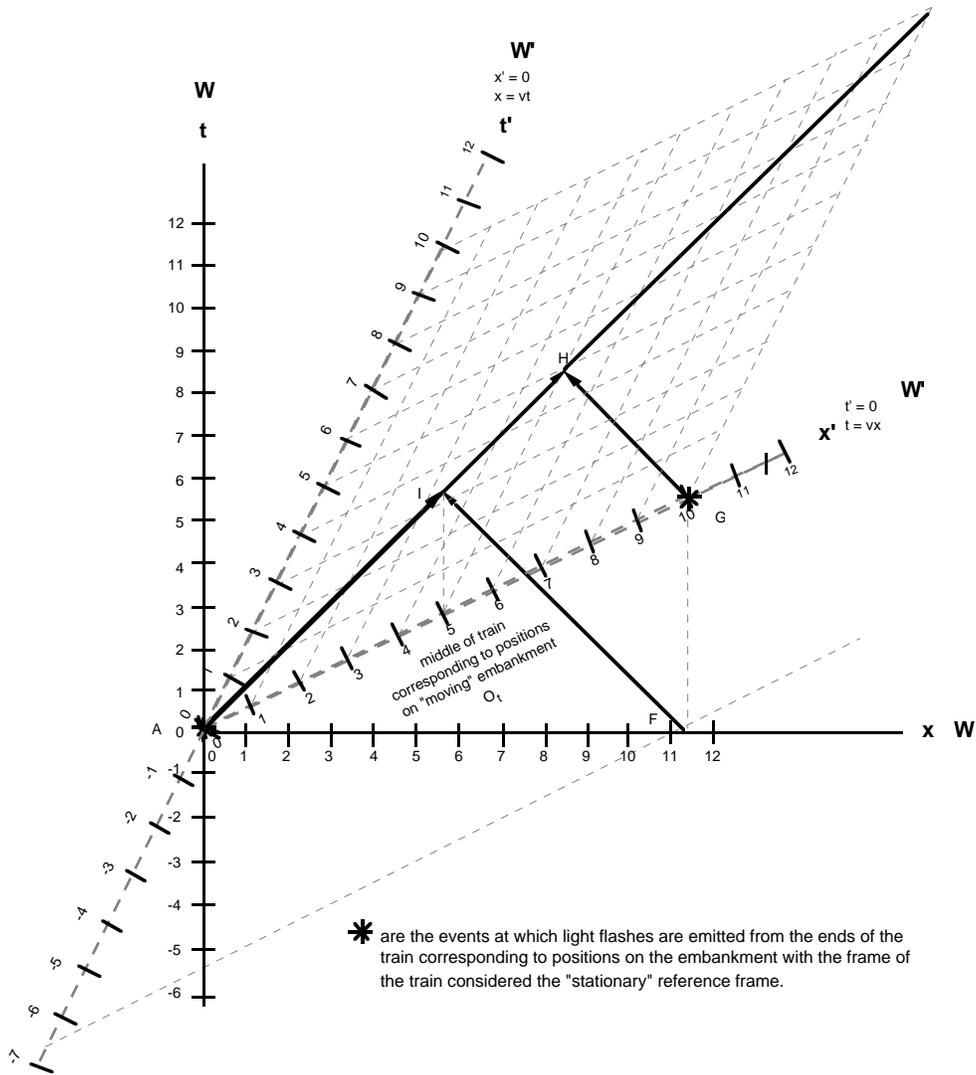

Figure 10. Minkowski diagram of W and W' showing simultaneity in the inertial reference frames of the train (W') and the embankment (W) with the train considered the "stationary" frame.

# On the Arbitrary Choice

"stationary" reference frame of the embankment now apply to the "stationary" reference frame of the train. One would think this applies to the world lines of the light flashes used to establish simultaneity in the "stationary" as well as the "moving" inertial reference frames. Regardless of the particular world lines of the light flashes, the distance between the flashes in whichever inertial reference frame is designated the "stationary" frame needs to be the same. In the reverse scenario, the distance between the light flashes used to establish simultaneity in the "stationary" inertial frame of the *train* is 10 units, the length of the train.

One can see in Figure 10, essentially Figure 9 with coordinates for the frame of the embankment, the analogous circumstances to Figure 8, only which reference frame is "stationary" and which is "moving" have been reversed. When the embankment is the "moving" reference frame, the events A and I are simultaneous in the "moving" reference frame of the embankment because the light flashes AI and FI meet Einstein's criterion for simultaneity. Since A and I correspond to points of the embankment at the same time, the length from A to I is a measurement of the length of a section of the embankment for observers at rest in the "moving" reference frame of the embankment. This length, $10/(1 - v^2/c^2)^{1/2}$ units, is established in the argument on the relativity of simultaneity after length is established in the "stationary" frame of the train where light flashes occur in events A and G that are equidistant from the observer at rest in the "stationary" frame of the train.

This interesting point should be emphasized. The length of an existent at rest in an inertial reference frame as measured by observers at rest in that reference frame is dependent on whether the reference frame is considered the "stationary" or "moving" reference frame in the argument on the relativity of simultaneity. In Figure 9, it can be seen that the light flashes used to delineate simultaneity in the reference frame of the train when it is considered the "stationary" frame are from event A to event H and from event G to event H. (Event H is the meeting of the light flashes originating in events A and G midpoint on the train.) Also, the light flashes from event A to event C and from event B to event C found in Figure 8 do not represent light flashes meeting O midpoint on the embankment when it is the "moving" inertial reference frame. Rather, as seen in Figure 9, the lights flashes from event A to event I and from event F to event I represent two light flashes meeting at O midpoint on the embankment when the embankment is the "moving" inertial reference frame. In terms of the logical sequence in the argument on the relativity of simultaneity, the two light flashes from A to I and from F to I are used to establish

# On the Arbitrary Choice

simultaneity in the "moving" reference frame of the embankment <u>after</u> simultaneity is first established in the "stationary" reference frame of the train using light flashes from A to H and from G to H. In an inspection of Figure 10, one can see a bit more precisely that the light flashes striking the ends of the train and the corresponding points on the embankment and which are used to establish simultaneity in the "stationary" inertial reference frame of the train do not serve to establish simultaneity in the "moving" inertial reference frame of the embankment. In Figure 10, the t coordinate for A is 0, but the t coordinate for G is close to 5. One can also see from an inspection of Figure 10 that where the t' coordinate for A is 0, the t' coordinate for F is between -6 and -7. For A and F, $x = 0$, and A and F are simultaneous events in the "moving" inertial reference frame of the embankment.

What has occurred is that the direction in which the argument on the relativity of simultaneity is made has been reversed, and this reversal has changed the nature of space in a particular inertial reference frame such that the length of the concrete measuring rod at rest in this reference frame and measured by observers at rest in this reference frame is different depending on whether the inertial reference frame in which the rod is at rest is the "stationary" or the "moving" reference frame in the argument on the relativity of simultaneity. (The fundamental tenet is adjusted to: the description of physical phenomena must be the same for observers at rest in their inertial reference frames when these frames are the "stationary" reference frames in the argument on the relativity of simultaneity and also the description of physical phenomena must be the same for observers at rest in their inertial reference frames when these frames are the "moving" reference frames in this argument.)

There is nothing in the physical world that prompts an individual considering inertial reference frames in uniform translational motion relative to one another to choose a particular direction in which to argue the relativity of simultaneity. It should be remembered that the possibility of making the argument in either direction without any limitation by the physical world is central to retaining the central tenet of the special theory that there is no preferred inertial reference frame from which to describe physical phenomena. If there were some reason in the physical world to choose one direction over another, there would be a preferred inertial reference frame from which to describe physical phenomena. The choice in arguing the relativity of simultaneity is an arbitrary one, a free one, on the part of the person considering the relativity of simultaneity. This arbitrary choice on the part of the individual considering the relativity of simultaneity is reflected in the observers each being at rest in their respective inertial reference frames in uniform translational motion relative to one another and considering the inertial reference frame in which each is at rest the "stationary" reference frame and the other observer's inertial reference frame the "moving" reference frame.

One might think that the different scenarios regarding spatiotemporal relations can be distinguished by the concrete physical existence of the measuring rod in the case of spatial relations or clocks in the case of temporal relations. That is, it might be thought that even though the relativity of simultaneity can be argued in either direction, in practice, only one scenario applies at any one time and that this limitation is imposed by the concrete physical world. Essentially, the basis for this thought is that the physical mechanism of a clock or the physical structure of the rod are concrete, that is, they are at rest in only one inertial reference frame and that their basic constitution remains unchanged in whatever inertial reference frame from which they are considered. It has been shown that with regard to measuring rods, this is not the case.

More generally, the essence of the functioning of a clock in the special theory is the motion of light over a prescribed distance in both possible directions. It is not the concrete existence of a clock that is at the essence of the functioning of a clock in an inertial reference frame. This is the case because the synchronization of clocks is established in Einstein's 1905 and 1917 definitions of simultaneity (or the synchronization of clocks) by the motion of light over a prescribed distance in both possible directions.[15] This motion of light can be considered periodic motion, and it is explicitly the periodic motion of light between two spatial points A and B in an inertial reference frame that Einstein relied upon in 1905 in defining simultaneity in his original paper on the special theory. Because the length of an existent in an inertial reference frame depends on determining the ends of the existent simultaneously, one would expect that the concrete nature of the existent would not inhibit the fundamental character of the relativity of simultaneity.

*8.2     The Time of the "Stationary" Inertial Reference Frame*

In the argument on the relativity of simultaneity, the "moving" observer relies on simultaneity and time of the "stationary" frame to determine that the criterion for simultaneity is not met by clocks at rest in the "moving" frame because the "moving" observer relies on clocks synchronized in the "stationary"

# On the Arbitrary Choice

frame to set clocks at rest in the "moving" frame and thus to determine whether the criterion for simultaneity is met in the "moving" frame. This has been shown in the train gedankenexperiment and will be shown in Einstein's original argument on the relativity on simultaneity. Einstein explicitly proposed in 1905 in his argument on the relativity of simultaneity that the clocks of the "moving" frame are synchronized in accordance with, and rely on, the clocks of the "stationary" frame. (It should be noted with regard to the train gedankenexperiment that if the "moving" observer synchronized his own clocks without relying on the synchronization of the clocks in the "stationary" inertial frame, the clocks of the "moving" observer would not be subject to the effective velocities c + v or c - v. Instead, they only consider the light flashes in terms of their finite and invariant velocity c.)

In his original paper on the special theory of relativity, Einstein noted that simultaneity (or the common time of clocks) is delineated for an inertial frame of reference when, by definition, the time required for a ray of light to travel from a spatial point A to a spatial point B is equal to the time required for a ray of light to travel from point B to point A.(16) He argued the relativity of simultaneity this way:

> We imagine further that at the two ends A and B of the rod [moving with uniform translational velocity relative to the stationary inertial system], clocks are placed which synchronize with the clocks of the stationary system, that is to say that their indications correspond at any instant to the "time of the stationary system" at the places where they happen to be. These clocks are therefore "synchronous in the stationary system."
>
> We imagine further that with each clock there is a moving observer, and that these observers apply to both clocks the criterion established...for the synchronization of two clocks [that the flight time of a light ray in an inertial frame of reference from spatial point A to spatial point B is equal to the flight time of a light ray from point B to point A]. Let a ray of light depart from A at the time* $t_A$, let it be reflected at B at the time $t_B$, and reach A again at the time $t'_A$. Taking into consideration the principle of the constancy of the velocity of light we find that

$$t_B - t_A = \frac{r_{AB}}{c - v} \quad \text{and} \quad t'_A - t_B = \frac{r_{AB}}{c + v}$$

> where $r_{AB}$ denotes the length of the moving rod--measured in the stationary system. Observers moving with the rod would thus find that the two clocks were not synchronous, while observers in the stationary system would declare the clocks to be synchronous.
> 
> \* Time here denotes "time of the stationary system" and also "position of hands of the moving clocks situated at the place under discussion."[17]

Whether using readings from clocks synchronized in the "stationary" frame or the coincidence of flashes of light as judged from the railway embankment, the "moving" observer relies on the time established first in the "stationary" frame in the argument on the relativity of simultaneity. This reliance by the "moving" observer on the time established in the "stationary" reference frame sets up a *different basis* for the development of simultaneity, and thus time, in his own frame than would otherwise have been the case if the "moving" observer had instead been the observer at rest in the "stationary" reference frame in the argument on the relativity of simultaneity.

*8.3    The Rate of Clocks at Rest in the "Stationary" and "Moving" Reference Frames*

It has been shown that the spatial length of a physical existent at rest in an inertial reference frame and aligned along the axis of uniform translational motion of two inertial reference frames relative to one another depends on whether the reference frame in which the rod is at rest is designated the "stationary" or "moving" reference frame in the argument on the relativity of simultaneity. A similar dependency holds for the rate of a clock in an inertial reference frame. The rate of a clock at rest in an inertial reference frame depends on whether the reference frame in which the clock is at rest is designated the "stationary" or "moving" reference frame in the argument on the relativity of simultaneity. In Figures 11 and 12, there are the analogous conditions for temporal durations in inertial reference frames in uniform translational motion relative to one another to those used to explore spatial length. In Figures 11 and 12, Einstein's original argument on the relativity of simultaneity is depicted in its essential elements. Allow that a distance of

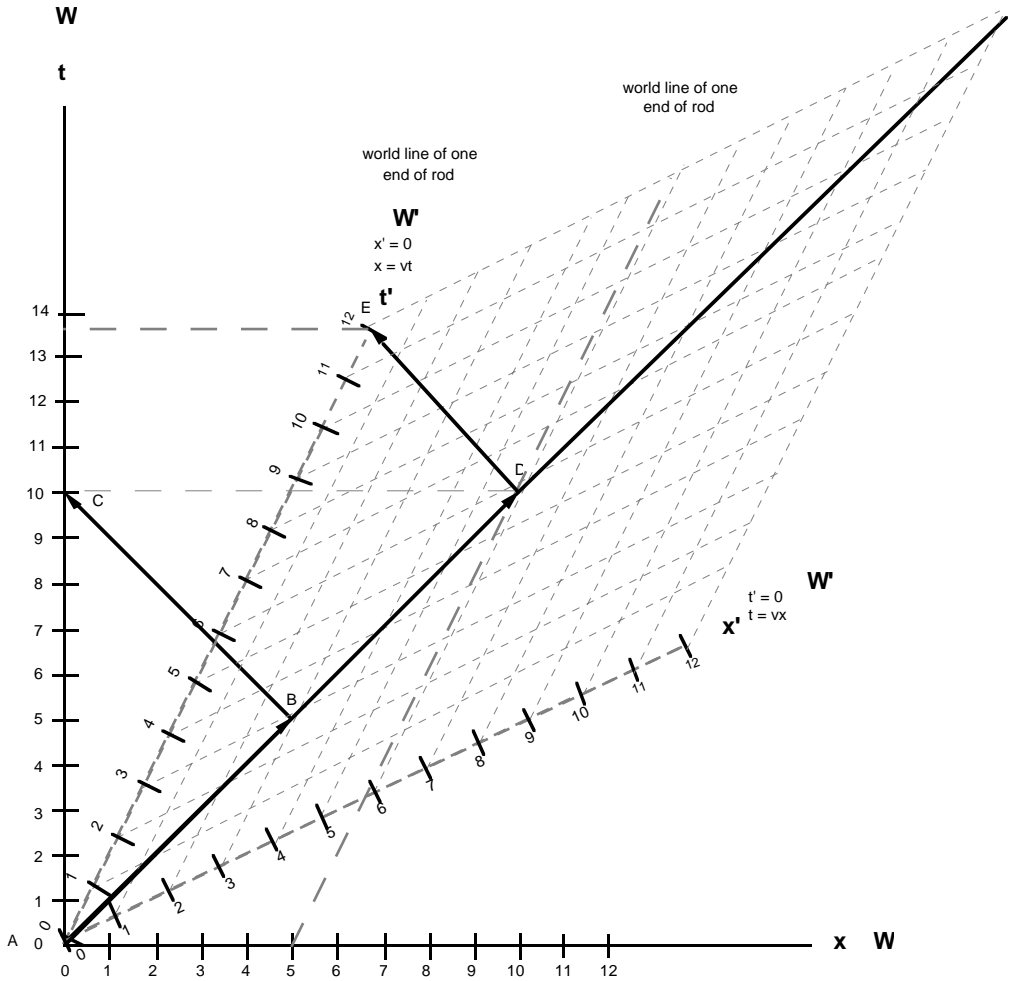

Figure 11. Minkowski diagram of inertial reference frames W and W' showing simultaneity in accordance with Einstein's 1905 formulation with W the "stationary" reference frame and W' the "moving" reference frame.

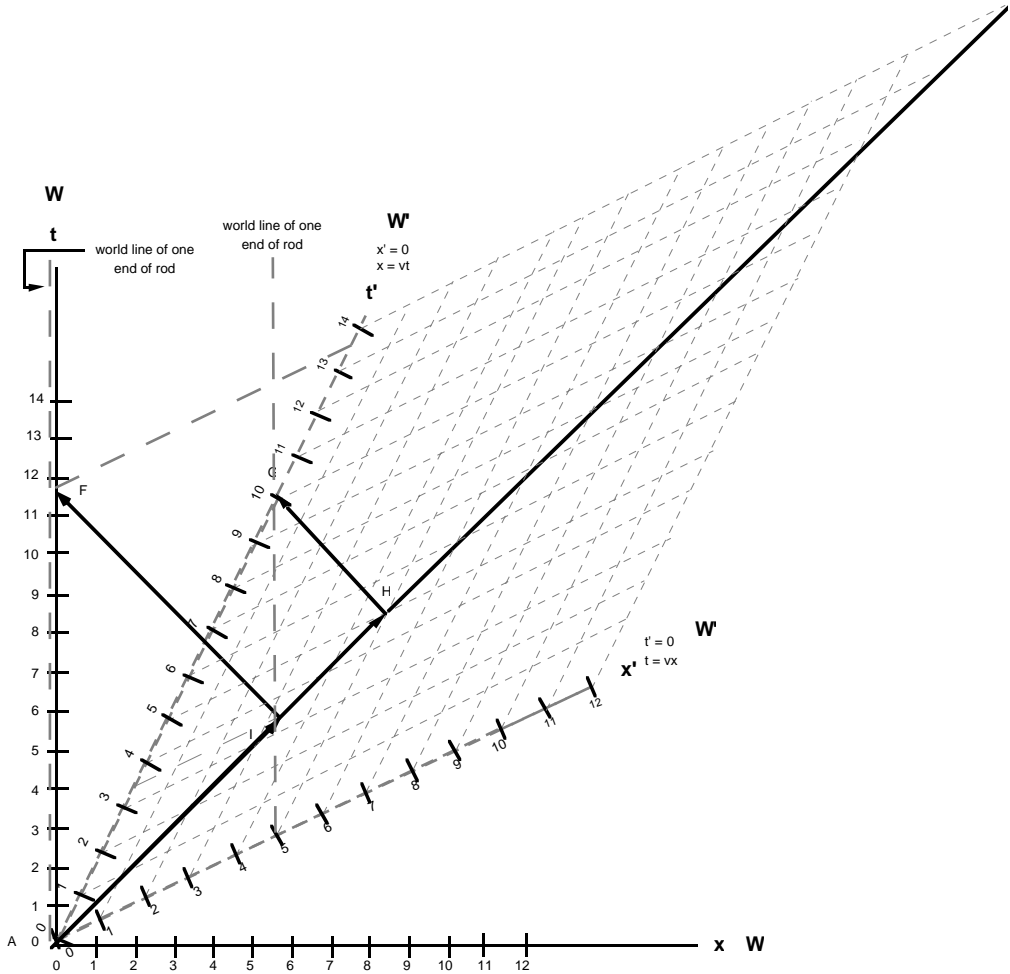

Figure 12. Minkowski diagram of inertial reference frames W and W' showing simultaneity in accordance with Einstein's 1905 formulation with W' the "stationary" reference frame and W the "moving" reference frame.

# On the Arbitrary Choice

Δx = 5 units is used to establish simultaneity in accord with Einstein's original criterion. If a light flash travels from event A to event B, traveling Δx, and if the light flash is reflected back without delay, it requires the same amount of time for the light flash to return to x = 0 from x = 5 (in event C) as it took for the light flash to travel from event A to event B. Einstein's original criterion for simultaneity in an inertial reference frame is met and time is established in W, the "stationary" frame. Time in W is represented by the t axis. It can be seen in Figure 11 that the simultaneity established in this process in W is not the same as the simultaneity established in W'. An inspection of this figure indicates that simultaneity in W' requires another process. Allow that a rod is at rest in W' and thus moving in a uniform translational manner relative to W and that its length as measured in W is 5 units. Allow that in line with Einstein's criterion, a light flash begins in event A at one end of the rod, travels to event D where it reaches the other end of the rod, and is immediately reflected back. (Event D occurs on the world line of the end of the rod toward which the light flash travels after first being emitted.) The light flash returns to the end of the rod that it originated from in event E at x' = 0. In terms of the time of W, $t_{AD} \neq t_{DE}$ and thus Einstein's original criterion for establishing simultaneity is not met in W' when this process is considered in terms of the time of W. From Figure 11, it can also be seen that $t'_{AD} \approx t'_{DE} \approx 6$ units and that simultaneity, and thus time, is established in W' in accordance with Einstein's original criterion for simultaneity in an inertial reference frame.

In terms of the time of W', t', the passage of the light ray from one end of the rod to the other and back again occurs in about 12 units. If the concern is with the amount of time that elapses in W, the "stationary" reference frame, corresponding to these 12 units in W', the "moving" reference frame, the horizontal dashed line parallel to the space axis for W indicates that the corresponding time in W is between 13 and 14 units.

Figure 12 displays the reverse scenario where W' is the "stationary" inertial reference frame and W is the "moving" inertial reference frame. Allow that a distance of Δx' = 5 units is used to establish simultaneity in accord with Einstein's original criterion. If a light flash travels from event A to event H, that covers Δx', and if the light flash were reflected back without delay, it requires the same amount of time for the light flash to return to x' = 0 from x' = 5 (in event F) as it took for the light flash to travel from event A to event H. Einstein's original criterion for simultaneity in an inertial reference frame is met and time is established in W', the "stationary" frame. Time in W' is represented by the t' axis. It can be seen in Figure 12 that the simultaneity established in this process in W' is not the same as the simultaneity established in W. An inspection of this figure indicates that simultaneity in W requires another process. Allow that a rod is at rest in W and thus moving in a uniform translational manner relative to W' and that its length as measured in W' is 5 units. Allow that in line with Einstein's criterion, a light flash begins in event A at one end of the rod, travels to event I where it reaches the other end of the rod, and is immediately reflected back. (Event I occurs on the world line of the end of the rod toward which the light flash travels after first being emitted.) The light flash returns to the end of the rod that it originated from in event F at $x = 0$. In terms of the time of W', $t'_{AI} \neq t'_{IF}$ and thus Einstein's original criterion for establishing simultaneity is not met in W when this process is considered in terms of the time of W'. From Figure 12, it can also be seen that $t'_{AI} \approx t'_{IF} \approx 6$ units and that simultaneity, and thus time, is established in W in accordance with Einstein's original criterion for simultaneity in an inertial reference frame.

In terms of the time of W, t, the passage of the light ray from one end of the rod to the other and back again occurs in about 12 units of time. If the concern is with the amount of time that elapses in W', the "stationary" reference frame corresponding to these 12 units in W, the "moving" reference frame, the dashed line parallel to the space axis for W' indicates that the corresponding time in W' is between 13 and 14 units.

Nothing physical is presented to distinguish the different scenarios in the argument on the relativity of simultaneity. All that distinguishes them is an arbitrary choice on the part of the individual arguing the relativity of simultaneity concerning the direction in which the argument is made. This arbitrary choice is anchored in the experience of observers that they are at rest in their respective inertial reference frames in uniform translational motion relative to one another and that each of their respective inertial reference frames is for them "stationary" while the other reference frame is "moving." This arbitrary choice concerning the direction in which to argue the relativity of simultaneity is made implicitly when an individual is concerned with other results of the special theory, for example the temporal duration of occurrences or the spatial length of physical existents in inertial reference frames in uniform translational motion relative to one another. And other results of the special theory depend on these spatial and temporal relations.

# On the Arbitrary Choice

### 9.   *An Interesting Circumstance*

There is an interesting circumstance that results from the possibility of arguing the relativity of simultaneity in either direction, that is with either of two inertial reference frames in uniform translational motion relative to one another designated the "stationary" frame while the other reference frame is designated the "moving" frame. This circumstance concerns the nature of the light flashes used to establish simultaneity in the inertial reference frames, in the case of the train gedankenexperiment the embankment and the train. This circumstance serves as an additional indication that the arbitrary decision as to which direction to argue the relativity of simultaneity has an impact on the physical world itself through affecting the light flashes seen by observers at rest in their inertial reference frames in establishing simultaneity in their respective inertial reference frames. The argument on the relativity of simultaneity affects the course in the physical world of these light flashes. This effect occurs even though the physical existence of the light flashes themselves appears not to be impacted. It is the same light flashes that are used in either scenario chosen.

There is a set of four world lines each (or, perhaps, more precisely, three world lines of which one has a shorter element) for defining simultaneity in the inertial reference frames of the embankment and the train in each scenario. There are two distinct sets, one for each scenario. The scenario in which the chosen direction has the reference frame of the embankment as the "stationary" inertial reference frame and the train as the "moving" inertial reference frame is depicted in Figure 8. In this scenario, the light flash emitted in event A and meeting in event C the observer at rest midway on the section of embankment the end positions of which correspond to the ends of the "moving" train, and the light flash emitted in event B and meeting the light flash from A in event C are used to establish simultaneity in the "stationary" inertial reference frame of the embankment. The light flash emitted in event A that meets the observer at rest on the embankment in event C and travels to event D where it meets the observer at rest midpoint on the train, and the light flash emitted in event E, from the other end of the train that meets the light flash from A in event D are used to establish simultaneity in the "moving" inertial reference frame of the train.

The reverse circumstance can be seen in Figure 10. Where the the inertial reference frame of the train is the "stationary" frame, the light flash emitted in event A and meeting in event H the observer at rest midway on the train the ends of which correspond to the positions on the "moving"

# On the Arbitrary Choice

embankment where the lightning flashes struck, and the light flash emitted in event G and meeting the light flash from A in event H are used to establish simultaneity in the "stationary" inertial reference frame of the train. The light flash emitted in event A that meets the observer at rest midpoint on the embankment in event I, and the light flash emitted in event F, from the other end of the embankment where lightning struck that meets the light flash from A in event I are used to establish simultaneity in the "moving" inertial reference frame of the embankment. (Because the embankment is "moving" in the direction of decreasing values of x' and thus towards the light flash emitted in event A, this light flash, in terms of the time of the "stationary" reference frame of the train, meets the observer midpoint on the embankment in event I before the light flash reaches the observer at rest midpoint on the train in event H.)

It has been shown that it is an arbitrary decision which frame is considered the "stationary" reference frame and which the "moving" reference frame in Einstein's argument on the relativity of simultaneity. Thus, the question arises: What is the relationship between the light flashes from A to C in Figure 9 and from A to H in Figure 10, as well as those light flashes from B to C in Figure 9 and from G to H in Figure 10, that is between the light flashes that are used to establish simultaneity in the reference frames of the embankment and train respectively when each is considered the "stationary" inertial reference frame in the argument on the relativity of simultaneity? The light flashes from A to C and from B to C can have logical priority in the argument, or the light flashes from B to C and from G to H can have logical priority. In terms of the argument, there is only this logical distinction, arbitrarily decided, that determines within each scenario which of the two different pairs of world lines will meet first and which will meet second.

How is it that there are two different sets of world lines in Figures 8 and 10 for the light flashes used to establish simultaneity in the inertial reference frame designated the "stationary" reference frame in the argument on the relativity of simultaneity? Are the light flashes from A to C and from A to H the same light flash? Or are the light flashes from B to C and from G to H the same light flash? For that matter are the light flashes from A to D in Figure 8 and from A to I in Figure 10 the same? And are the light flashes E to D in Figure 8 and from F to I in Figure 10 the same? As can be seen by their world lines in Figures 8 and 10, the members of each pair appear to be different flashes of light in the context of their roles in one of the scenarios in the argument on the relativity of simultaneity (i.e., with one inertial reference frame designated the "stationary" reference frame and the other inertial reference frame designated the "moving" reference frame).  But since the argument on the relativity of simultaneity can be made in either direction without physical constraint and thus with the same light flashes, it would seem that the light flashes from A to C and from A to H are indeed the same light flash, the light flashes from B to C and from G to H are the same light flash, the light flashes from A to D and from A to I are the same light flash, and the light flashes from E to D and from F to I are the same light flash.  One might expect that if this is the case, instead of two world lines for each of the light flashes that play the same role in the argument on the relativity of simultaneity, albeit in different scenarios, there would be only one world line for the light flashes.  Instead, the argument on the relativity of simultaneity indicates that a light flash involved in establishing simultaneity in an inertial reference frame has two world lines in the special theory.

10.     *Whose Clocks Run Slower?*

The importance of an observer's considering himself at rest in an inertial reference frame and the observer's considering this reference frame at rest, or the "stationary" reference frame, to the special theory has been discussed.  Essentially, if it were not the case for all observers in inertial reference frames in uniform translational motion relative to one another, there would exist some distinction between the observers' experiences of motion in their respective inertial reference frames.  The relativity of simultaneity, as well as the Lorentz coordinate transformation equations, then could not be properly derived.  A preferred inertial reference frame would exist that would provide the basis for the spatial and temporal structure for the description of events in the physical world.  Also, if there were some difference in the motion experienced by observers at rest in their respective inertial reference frames, then the relativity of simultaneity and the Lorentz coordinate transformation equations could not be derived with either inertial reference frame being the "stationary" reference frame and the other reference frame being the "moving" reference frame.

The central role of the arbitrary choice in the argument on the relativity of simultaneity concerning which inertial reference frame is "stationary" and which "moving" in the special theory has also been discussed.  Among the consequences are different forms of the Lorentz coordinate transformation equations and different equations for relating the spatial length of physical existents along the axis of relative motion and the duration of occurrences between two inertial reference frames moving in a uniform translational manner relative to one another.  Unlike the twin paradox where physicists generally

## On the Arbitrary Choice

maintain that one twin is not always in an inertial reference frame, in the problem at hand, observers are always at rest in their respective inertial reference frames. Thus, the question remains, whose clock runs slower?

This question becomes particularly important because it can serve to emphasize that *the exact same physical circumstances* can provide the physical basis for *different empirical results*, for example whose clocks run slower. In one scenario, an observer A, who is in the "moving" inertial reference frame, has the clocks that run slower in comparison with the clocks that an observer B has in the "stationary" reference frame. In the other scenario, B's clocks run slower than A's clocks. As we have seen, nothing in the physical foundation of these scenarios need be different. Only the direction in which the relativity of simultaneity is argued is different (which inertial reference frame is the "stationary" frame and which the "moving" frame), a difference that depends upon, and helps to explain the significance of, the experience of each of the observers being at rest in the their respective inertial reference frames and considering their respective frames the "stationary" frame.

Indeed, if only one scenario held, there would be a preferred reference frame, the reference frame in this scenario from which the other inertial reference frame is judged to be "moving." This circumstance would violate the fundamental tenet of the special theory that inertial reference frames in uniform motion relative to one another are equivalent for the description of physical phenomena. What then is behind the different empirical results if it is not something physical? Whatever it is, it must account first and foremost for the different directions in arguing the relativity of simultaneity that is the foundation for the other results of the special theory, including the relationship between temporal durations and spatial distances between inertial reference frames in uniform translational motion relative to one another. It is most likely that it is some cognitive factor. The first reason is that the distinction in direction is on the level of the argument itself. The second reason is that there is a decision made by an individual considering the physical circumstances concerning in which direction the relativity of simultaneity is to be argued, which inertial frame is the "stationary" reference frame in which simultaneity is first defined and which the "moving" reference frame.

This decision is essentially a psychological act, one which has distinct consequences in the physical world depending on the decision made, namely distinct sets of empirical results that support the special theory, including the temporal and spatial relationships between inertial reference frames in uniform translational motion relative to one another. Concerning the distinct sets of results, each is just as likely because the argument on the relativity of simultaneity can be argued in either direction without physical or mental constraint. Either of the two inertial reference frames in uniform translational motion relative to one another can be designated the "stationary" frame while the other frame is then designated the "moving" frame. The decision to argue in one direction or the other direction is truly a free one on the part of the individual making the argument and because this decision concerns an *argument* while concrete physical circumstances can support either direction in which the argument may be made without prompting a decision in one direction, the decision as to which direction to argue the relativity is cognitive in nature, not reducible to a physical substrate, and impacts the physical world.

As noted, the arbitrary choice concerning the direction in arguing the relativity of simultaneity extends to the problem raised at the beginning of this paper, namely the relation of temporal durations of an occurrence in inertial reference frames in uniform translational motion relative to one another. In this case, the arbitrary decision as to the direction in which to argue the relativity of simultaneity leads to the question concerning which clocks at rest in inertial reference frames in uniform translational motion relative to one another run slower? According to the analysis presented here, one would expect that each observer at rest in his inertial reference frame and considering his frame "stationary" and the other frame "moving" would find that the other observer's clocks are running slower by exactly the same amount. Thus, one should find in experiments where observers at rest in their respective inertial reference frames in uniform translational motion relative to one another determine the rate of physical processes occurring in one location in the other inertial reference frame (e.g., the rate of a clock at rest in this other inertial reference frame) that both observers will record the same slowing of time in the other inertial reference frame. This condition is a test of the relationship of temporal duration for some occurrence in both directions and is thus a test of the possibility of arguing the relativity of simultaneity in both directions while the concrete experimental circumstances in the physical world are the same in both directions.

Experiments, though, have not in general tested this condition. Instead, experiments testing the equations derived in the special theory concerning spacetime that relate the durations of occurrences or the length of physical existents in inertial reference frames in uniform motion relative to one another have generally met one of two other conditions:

> Condition 1: One frame is not always inertial in nature (it is at some point in its motion accelerating);
>
> Condition 2: The test is made in only one direction, that is with one inertial frame considered the "stationary" frame and the other inertial reference frame considered the "moving" reference frame.

Examples of both conditions will be discussed. This discussion of conditions 1 and 2 will demonstrate the significant attention devoted to them and place in contrast the relative lack of attention paid to empirically testing the relationship between temporal duration in inertial reference frames moving in uniform translational motion relative to one another where each of the inertial reference frames in one of the two scenarios is the "stationary" frame while the frame not selected to be the "stationary" frame is the "moving" frame.

*10.1  Clocks Flown Around the World*

Hafele and Keating reported an experiment that attempted to take into account a key aspect of the twin paradox gedankenexperiment, namely where one reference frame is always considered inertial and the other cannot be considered inertial in at least part of its motion.[18,19] This experiment was a test that met condition 1. Hafele and Keating compared the time kept by the atomic time scale of the U.S. Naval Observatory with the time kept by four cesium clocks that were flown around the world in jet aircraft. In one trip, the clocks were flown eastward around the world, and in another trip, they were flown westward around the world. On each trip, the clocks were thus involved in a round trip, which is a central aspect of the twin paradox. Hafele and Keating introduced a hypothetical inertial reference frame that was, of course, not rotating.

Using the special theory, specifically variations of equations 1 and 2 (the equations that relate the durations of an occurrence in inertial reference frames in uniform translational motion relative to one another), as well as taking into account other considerations specific to their experiment including a general relativistic consideration, Hafele and Keating developed predictions concerning time differences between the time kept by the U.S. Naval Observatory and the four cesium clocks on their two trips. They considered both the U.S. Naval Observatory as well as the clocks circumnavigating the earth as part of rotating reference frames relative to the hypothetical inertial reference frame. Though

## On the Arbitrary Choice

these rotating reference frames experienced acceleration, Hafele and Keating maintained that the rotating reference frames could, in essence, constitute inertial reference frames traveling in a uniform translational manner relative to a hypothetical inertial reference frame. Thus, the relation between the duration of an occurrence in inertial reference frames in uniform translational motion relative to one another would be applicable to these rotating reference frames. Essentially, they considered the hypothetical inertial reference frame the "stationary" reference frame and the rotating reference frames as "moving" reference frames. The experimental results confirmed their predictions.

It should be emphasized, though, that their experiment concerned the twin paradox gedankenexperiment specifically and not the type of circumstance of concern here. It should also be emphasized that this was an indirect test as clocks in the hypothetical inertial reference frame were not involved in the experimental measurement. Both the clocks in their circumnavigation of the world and the U.S. Naval Observatory were in accelerating reference frames, reference frames for which the special theory, in general, does not provide an adequate description. The central point for Hafele and Keating though, is that as noted, the twin paradox involves one twin changing direction during his travel and the return of this twin to the other twin. And for Hafele and Keating, the clocks on the airplanes as well as the atomic time scale of the U.S. Naval Observatory are considered to have taken a roundtrip relative to the hypothetical clocks in the inertial reference frame that have remained "stationary."

### 10.2  *The Lifetime of μ-mesons*

Consider condition 2 where the the experiment has been conducted with only one of the inertial reference frames considered the "stationary" frame and the other inertial reference frame considered the "moving" frame. Following work by Rossi and Hall[20], Frisch and Smith[21] used the distribution of μ-meson decay as a clock.

> As far as we know the probability of the radioactive decay of subatomic particles, and thus the average time they survive before decaying, is set by forces entirely internal to their structure. Therefore, any dependence of the decay probability of radioactive particles on their speed is an example of a general property of clocks in motion relative to an observer.[22]

Specifically, Frisch and Smith determined the decay distribution over time for μ-mesons in the earth's atmosphere after they are brought to rest in an

## On the Arbitrary Choice

inertial reference frame. Using this decay distribution, Frisch and Smith developed predictions both with and without time dilation concerning the number of μ-mesons traveling at an average velocity of 0.99c (where c is the velocity of light) that would survive a journey of 1907 meters downward through the atmosphere to near sea level. In one experimental trial, for example, instead of the 27 μ-mesons predicted to survive the journey out of a total of 568 μ-mesons if there were no time dilation, 412 μ-mesons survived the journey. Using the μ-meson decay distribution as a clock, this number of surviving μ-mesons indicated an elapsed time of 0.7 μsec. as compared to an elapsed time of 6.4 μsec., which is the quotient of the 1907 meters traveled divided by the velocity of the μ-mesons (i.e., in this case 0.9552c). The observed result indicated that the μ-meson clocks were "*running slow by a factor of about nine*" relative to the laboratory reference frame.[23] Various possible problematic factors, such as the deceleration of the μ-mesons through the atmosphere, were well accounted for by Frisch and Smith. Using their data in a different manner to that described above, Frisch and Smith derived a more accurate value for the time dilation in their experiment and found an observed time dilation factor of $8.8 \pm 0.8$. The predicted time dilation factor based on the special theory for their experiment was $1/(1 - v^2/c^2)^{1/2} = 8.4 \pm 2$.

Though it sounds odd to mention, Frisch and Smith did not address the point that theoretically the μ-mesons could be considered to have been at rest in "stationary" inertial reference frames and that the laboratory reference frame is the "moving" inertial reference frame. In this scenario, the laboratory reference frame, including the people in it, would be aging very slowly, roughly 8.5 times slower than when the laboratory reference frame on earth is considered the "stationary" reference frame at rest. If equation 1 is used to describe the relation between the temporal durations in the "stationary" frame of the laboratory reference frame and the "moving" frame of the μ-mesons, equation 2 would be the appropriate equation for relating temporal durations in these inertial reference frames in the reverse scenario that Frisch and Smith did not consider. In their discussion of the μ-meson decay experiment, Feynman, Leighton, and Sands[24] also did not discuss the theoretical possibility of an alternative scenario along the lines indicated above.

French[25] did consider that the reciprocal scenario is theoretically possible. He did so in order to demonstrate that the number of decays in both scenarios would need to be the same. French wanted to uphold the tenet that physical description for observers at rest in their respective inertial reference frames in uniform translational motion relative to one another is equivalent. French did not consider the reciprocal scenario in order to emphasize the point that slower aging in the laboratory reference frame on earth (when it is considered the "moving" reference frame) is just as likely as slower aging, and thus rate of decay, of the μ-mesons when the laboratory reference frame is considered the "stationary" reference frame.

*10.3    A Test in Both Directions*

It is predicted that, in accordance with the special theory, if a test were made concerning the relative rate of clocks in two inertial reference frames in uniform translational motion relative to one another, one would find that the rate of the clock at rest in one inertial reference frame is slower to observers at rest in the other inertial reference frame by exactly the same amount as a clock at rest in this latter frame is slower to observers at rest in the former inertial frame. Observers at rest in an inertial frame A would find clocks at rest in inertial frame B, moving in a uniform translational manner relative to A, run slow relative to clocks in A by exactly the same amount that clocks at rest in B run slow for observers at rest in A.  It seems that if the ingenuity that has been devoted to developing tests of the twin paradox or time dilation in one direction were directed toward developing empirical tests of time dilation in <u>both</u> directions (i.e., in which each of the inertial reference frames is in one scenario the "stationary" reference frame), empirical tests could be developed.  One specific possibility will be discussed shortly.

This prediction of time dilation in both directions is the basis for an empirical test of the relationship of temporal durations for some occurrence in inertial reference frames in uniform translational motion relative to one another in both directions.  It is the basis for a test of the possibility of arguing the relativity of simultaneity in both directions while the concrete experimental circumstances in the physical world are the same.

Here is one scenario that would test this prediction concerning the relative rates of clocks in inertial reference frames in uniform translational motion relative to one another.  The test is indirect but, nonetheless, conclusive.  This scenario relies on the Doppler effect in the special theory.

Allow that two inertial reference frames (W and W') with observers at rest in each inertial frame (O and O', respectively) move away from one another in a uniform translational manner relative to one another.  Let there be identically constructed cesium clocks at rest in one or the other of these inertial reference frames. Whenever a specific unit of time is measured by the clocks at rest in their respective inertial frames (for example, one hour), let a light pulse be emitted in the direction of the other inertial reference frame along the axis of uniform translational motion of the inertial reference frames relative to one another. The longitudinal Doppler effect in the special theory predicts that the relation of the frequencies measured by the observers at rest in their respective inertial reference frames has two possible forms. One form of the longitudinal Doppler effect is:

$$f' = f ((c - v)/(c + v))^{1/2} \quad (11)$$

where c is the invariant velocity of light in inertial reference frames, v is the uniform translational velocity of the inertial reference frames relative to one another, f' is the frequency of the light pulses measured by O' at rest in W' that emanate from the clock at rest in W, and f is the frequency of the light pulses from the clock at rest in W determined by O at rest in W.

The other form for the longitudinal Doppler effect is:

$$f = f' ((c - v)/(c + v))^{1/2} \quad (12)$$

where f is the frequency of the light pulses measured by O at rest in W that emanate from the clock at rest in W', and f' is the frequency of the light pulses from the clock at rest in W' determined by O' at rest in W'.

The longitudinal Doppler effect in the special theory is reciprocal. That the relationship in the special theory between temporal durations of an occurrence in inertial reference frames in uniform translational motion relative to one another is a factor in the longitudinal Doppler effect can be seen in the general formulation of the Doppler effect in the special theory. One version of this general formulation is:

$$f' = (f (1 - v^2/c^2)^{1/2})/(1 - (v/c)(\cos \Theta)) \quad (13)$$

where the light pulses emanating from the clock at rest in W are being considered and cos is the angle of the light ray in W relative to the axis along which W and W' are in uniform translational motion relative to one another. In the case of the longitudinal Doppler effect where the clock is at rest in W and the observer at rest in W' are moving away from one another, $\Theta = \pi$ and $\cos \Theta = -1$. Thus:

$$f' = (f (1 - v^2/c^2)^{1/2})/(1 + v/c) \quad (14)$$

$$f' = (f (1 + v/c)^{1/2} (1 - v/c)^{1/2})/(1 + v/c)$$

On the Arbitrary Choice

$$f' = (f (1 - v/c)^{1/2})/(1 + v/c)^{1/2}$$

$$f' = f ((c - v)/(c + v))^{1/2} \quad (11)$$

One finds the term, $(1 - v^2/c^2)^{1/2}$, in equation 13, a term that is also found in equations 1 and 2 relating the temporal durations of an occurrence in inertial reference frames in uniform translational motion relative to one another. Where $\cos = /2$, equation 13 becomes:

$$f' = f (1 - v^2/c^2)^{1/2} \quad (15)$$

and the time dilation effect becomes more apparent in what is known as the transverse Doppler effect. That is, the slower rate of the "moving" clock in W' by the factor $(1 - v^2/c^2)^{1/2}$ accounts for the difference in frequencies of the light pulses emitted by the clock at rest in W for O at rest in W and O' at rest in W'.

Where $v \ll c$, the factor $(1 - v^2/c^2)^{1/2}$ is very close to 1 and time dilation can be considered a negligible factor in the Doppler effect. Thus:

$$f' = (f (1 - v^2/c^2)^{1/2})/(1 - (v/c)(\cos \Theta)) \quad (13)$$

$$f' \approx (f)/(1 - (v/c)(\cos \Theta))$$

$$f' \approx (f)(1 + (v/c)(\cos \Theta)) \quad (16)$$

Equation 16 takes a form of the non-relativistic Doppler effect.

Consider the following scenario that is adapted from a version of the twin paradox experiment proposed by Darwin[26] and discussed by Resnick.[27] Two spaceships (A and B) are at rest in space far from any large body. In each spaceship there is an observer at rest. Each spaceship has identically constructed cesium clocks. Let the spaceships move in a uniform translational manner relative to one another and apart from one another. Let each cesium clock emit a light pulse after an hour has elapsed according to the time of the inertial frame in which a clock is at rest. Then both equations 11 and 12 above will hold, depending on which observer at rest in his reference frame is measuring the frequency of the light pulses emitted from the cesium clock at rest in the other spaceship. Allow that the uniform translational velocity is 0.8c. Then equation 11 becomes:

$$f' = f ((c - 0.8c)/(c + 0.8c))^{1/2}$$

$$f' = f ((0.2c)/(1.8c))^{1/2}$$

$$f' = (0.111)^{1/2} f$$
$$f' = 0.333 f .$$

In a similar manner, equation 12 becomes:

$$f = f' ((c - 0.8c)/(c + 0.8c))^{1/2} ,$$

and one can derive that:

$$f = 0.333 f' .$$

When an observer in one of the spaceships measures the frequency of pulses emitted by the clock at rest in the other spaceship, the frequency of the pulses will be 0.333 that found by the observer in the other spaceship when this latter observer measures the frequency of the light pulses emitted by the clock at rest in his spaceship. This is what is indicated in Figure 13 over three unit time periods for observers in both spaceships as measured by clocks at rest in their respective inertial reference frames. Thus the observer in spaceship B receives a light pulse from spaceship A once every three units of time as time is measured in B while A measures the frequency of the light pulses emitted by the clock at rest in A as one pulse per unit of time. In a reciprocal manner, the observer in spaceship A receives a light pulse from spaceship B once every three units of time as time is measured in A while B measures the frequency of the light pulses emitted by the clock at rest in B as one pulse per unit of time. The frequency of the light pulses measured by an observer at rest in one of the spaceships that emanate from the clock at rest in the other spaceship will also be 0.333 the rate of the pulses emitted by the identically constructed clock at rest in his own inertial reference frame. It is this clock at rest in his own inertial reference frame that the observer uses to determine the frequency of the light pulses from the other clock.

Only a part of the longitudinal Doppler effect is due to time dilation. The time dilation factor in equation 13 is $(1 - v^2/c^2)^{1/2}$. Given a uniform translational velocity of 0.8c:

$$(1 - v^2/c^2)^{1/2} = (1 - (0.8c)^2/c^2))^{1/2}$$
$$(1 - v^2/c^2)^{1/2} = (1 - (0.64c^2)/(c^2))^{1/2}$$
$$(1 - v^2/c^2)^{1/2} = 0.6 .$$

If time dilation did not effect the special relativistic Doppler effect in the proposed gedankenexperiment:

$$f' = (f (1 - v^2/c^2)^{1/2})/(1 + v/c) \quad (14)$$

# On the Arbitrary Choice

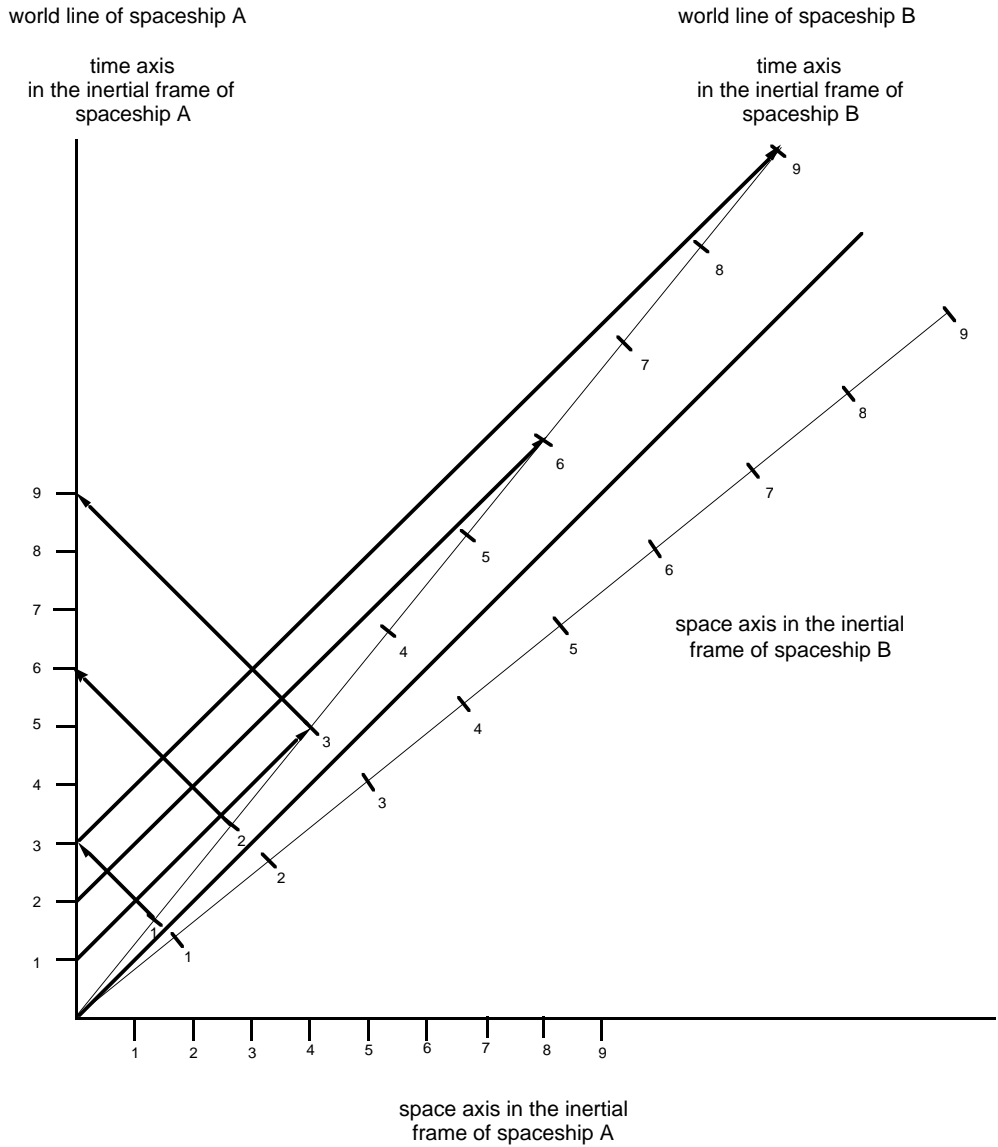

Figure 13. Spacetime diagram of world lines of light pulses from cesium clocks at rest in spaceship A and spaceship B.

$$f' = f/(1 + (v/c)) \quad (17)$$
$$f' = f/(1 + ((0.8c)/(c)))$$
$$f' = f/1.8$$
$$f' = 0.555\ f\ .$$

Using $(1 - v^2/c^2)^{1/2} = 0.6$ in equation 14, one obtains:

$$f' = (0.6\ f)/(1.8)$$
$$f' = 0.333\ f\ .$$

One can see that the impact of time dilation on the longitudinal Doppler effect in the gedankenexperiment is significant.

At a uniform relative velocity of 10 km per sec, though, the contribution of time dilation to the longitudinal Doppler effect would be difficult, although perhaps possible, to detect. It is simpler to use the transverse Doppler effect (i.e., equation 15). This way any difference between the frequencies of the light pulses emitted by one of the cesium clocks at rest in its respective spaceship that are recorded by observers at rest in spaceships A and B would be due to time dilation. Instead of the light pulses traveling along the axis of uniform translational motion relative to one another, the light pulses would travel orthogonally to this axis in the reference frame in which the clock that emitted the light flash is at rest.

With a uniform translational velocity $v = 10 \times 10^3$ m/sec:

$$(1 - v^2/c^2)^{1/2} = (1 - (10 \times 10^3 \text{ m/sec})^2/(2.99792458 \times 10^8 \text{ m/sec})^2)^{1/2}$$

$$(1 - v^2/c^2)^{1/2} = (1 - (1.0 \times 10^8 \text{ m/sec})/(8.987551787 \times 10^{16} \text{ m/sec}))^{1/2}\ .$$

In a binomial expansion to second order:

$$(1 - v^2/c^2)^{1/2} = 1 - 5.563 \times 10^{-10}\ .$$

This value is very close to 1. Nonetheless, the impact of the difference between this value and 1, on the order of $10^{-10}$, indicates that the transverse Doppler effect in equation 15 is large enough so that the difference in frequencies can be detected with the use of cesium clocks.

In a suitable arrangement of the spaceships, observers at rest in each spaceship using cesium clocks at rest in their respective spaceships could both

## On the Arbitrary Choice

detect the transverse Doppler effect. These observers would find that the frequency of the light pulses emitted from the other "moving" spaceship would be less than the frequency of the light pulses emitted by the cesium clock at rest in their own spaceship. Because this transverse Doppler effect would depend only on time dilation, this result would confirm that time dilation occurs in a reciprocal manner for observers at rest in their respective inertial reference frames.

The distance between the spaceships along the orthogonal axis which the light pulses travel would need to be large enough to assure that the different frequencies measured by observers at rest in their respective spaceships were indeed reflecting only a transverse Doppler effect, not confounded by other wave effects (due to what will be called wave compression/expansion components). A large enough distance would allow for enough time between the change from the blue shift to the red shift as one spaceship passes the other along an orthogonal to the axis of uniform translational motion relative to one another to detect a transverse Doppler effect.

Following is a deduction of the required distance. Figure 14 presents one possible representation of the experiment where the observer at rest in spaceship A considers spaceship B to be in uniform translational motion with velocity v relative to spaceship A. Let the time $t = 0$ at the instant that spaceship B lies along an orthogonal axis to the axis of uniform translational motion of spaceships A and B relative to one another. In this case, $\Theta = \pi/2$. As shown in Figure 14, after spaceship B has traveled the distance $v\Delta t$ in the brief time $\Delta t$:

$$\alpha = \pi - \beta.$$

Then:

$$\alpha = \pi/2 - \beta$$
$$\alpha = \pi/2 - \pi + \Theta$$
$$\alpha = \Theta - \pi/2.$$

Rearranging terms, one obtains:

$$\Theta = \alpha + \pi/2.$$

Taking the cosine of both sides of the equation, one obtains:

$$\cos \Theta = \cos (\alpha + \pi/2)$$

- 59 -

On the Arbitrary Choice

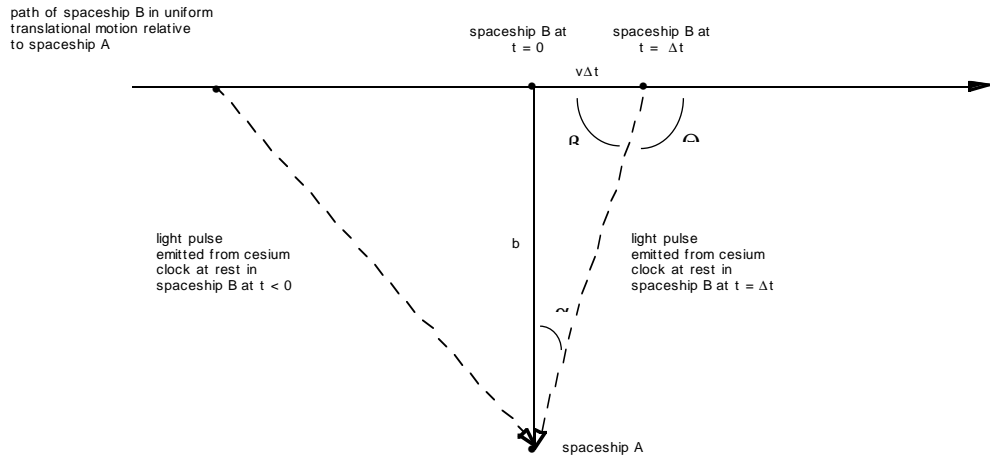

Figure 14. Transverse Doppler effect experiment where spaceship B is considered "moving" relative to spaceship A.

# On the Arbitrary Choice

$\cos \Theta = -\sin \alpha$.

$\alpha$ is small when $\Delta t$ is small. When $\alpha$ is small:

$\sin \alpha \approx \tan \alpha$

$\sin \alpha \approx v\Delta t/b$

where b is the distance along the orthogonal axis separating the spaceships.

Thus:

$\cos \Theta \approx -v\Delta t/b$.

The wave compression/expansion component of the special relativistic Doppler effect expressed in equation 13 is $1/(1 - (v/c)(\cos\ ))$. Thus for a small and a small amount of time $\Delta t$:

$1/(1 - (v/c)(\cos\ )) \approx 1/(1 + (v/c)(v\Delta t/b))$

$\approx 1 - (v/c)(v\Delta t/b)$.

The wave compression/expansion component for a small value of and $\Delta t$ is $v^2\Delta t/cb$.

As noted, the time dilation component in equation 13 is $(1 - v^2/c^2)^{1/2}$. With $v \ll c$, by the binomial expansion to second order:

$(1 - v^2/c^2)^{1/2} \approx 1 - 1/2\ v^2/c^2$

where the difference from 1 is $1/2\ v^2/c^2$. As can be seen from an inspection of equation 13, the time dilation component needs to be greater than the wave compression/expansion component in order to insure that the time dilation component is a major factor in the measured Doppler shift. Essentially, this means that the following relation should hold for the condition $v \ll c$:

$1/2\ v^2/c^2 > v^2\Delta t/cb$.

Then:

$b > 2c\Delta t$.

Since essentially the same analysis applies to the approach of spaceship B to an orthogonal axis with spaceship A, the measurement time $T = 2\Delta t$ or $\Delta t = T/2$. Then our result is:

$b > cT$.

If, for example, the experiment can be conducted in 1 second, with $T = 1$ sec, then b would need to be greater than 1 light-sec. If the experiment can be

## On the Arbitrary Choice

conducted in 0.001 sec, then b would need to be greater than 0.001 light-sec, or about 300 km.

*11.    Conclusion*

An exploration of a variation of the twin paradox leads to interesting results concerning the impact of the argument concerning the relativity of simultaneity in the special theory on the structure and functioning of the physical world. It has been shown that the same physical circumstances can support the relativity of simultaneity being argued in either of two directions with accompanying distinct sets of empirical results. The expression of this argument in the physical world relies upon an observer who considers himself at rest in an inertial reference frame and considers his or her reference frame "stationary," or at rest. This reliance extends to other results in the special theory because these other results depend on the relativity of simultaneity. The argument on the relativity of simultaneity also provides a rationale for an observer considering himself at rest in an inertial reference frame, this inertial reference frame in which he is at rest "stationary," and another inertial reference frame in uniform translational motion relative to him "moving."

Thus, in the special theory there is a role for a cognitive element influencing the structure, including the spatiotemporal structure, and functioning of the physical world. This conclusion is supported by the freely made decision noted in the special theory on the part of an individual as regards the direction in which to argue the relativity of simultaneity. Because this choice is not constrained by any physical factor in the special theory, it is also itself not reducible to any physical substrate.

In the special theory, it is predicted that observers at rest in their respective inertial reference frames which they consider to be "stationary," or at rest, and which are in uniform translational motion relative to one another, will find that the clocks at rest in the other observers' inertial reference frame will run slower and that they will run slower by exactly the same amount. This basic prediction of the special theory has not been subjected to the same degree of empirical scrutiny as the twin paradox or a one-way empirical test, such as the μ-meson experiment discussed.An interesting result concerning the light flashes used in establishing simultaneity in inertial reference frames in uniform translational motion relative to one another in a variation of Einstein's 1917 argument on the relativity of simultaneity was pointed out that serves to emphasize the consequences in the physical world of the particular direction that

## On the Arbitrary Choice

is arbitrarily chosen by an individual in arguing the relativity of simultaneity.

It is proposed that empirical tests in two directions (i.e., with each of the inertial reference frames the "stationary" reference frame in one scenario and the reference frame not designated the "stationary" reference frame is the "moving" reference frame) be developed and conducted. If empirical results from such tests support the predictions of the special theory, these empirical results would, in addition, support the thesis that there is a cognitive factor in in the argument on the relativity of simultaneity that affects results derived in the special theory concerning the structure and functioning of the physical world. It is expected that empirical tests in both directions concerning time dilation will support the predictions of the special theory. One such test is proposed.

# On the Arbitrary Choice

Acknowledgement

I thank Dr. Art Huffman for his significant contribution to the development of the proposed experiment involving the transverse Doppler effect in the special theory.

# On the Arbitrary Choice

Endnotes

[1]It appears Einstein maintained that the observer in the reference frame that at some point in its motion experienced acceleration would indeed have the clocks that ran more slowly. For example, in his original paper on the special theory, he first demonstrated the phenomenon of time dilation. He then wrote:

> From this [time dilation] there ensues the following peculiar consequence. If at the points A and B of K [the "stationary" reference frame] there are stationary clocks which, viewed in the stationary system, are synchronous; and if the clock at A is moved with the velocity v along the line AB to B, then on its arrival at B the two clocks no longer synchronize, but the clock moved from A to B lags behind the other which has remained at B by $(1/2)tv^2/c^2$ (up to magnitudes of fourth and higher order), t being the time occupied in the journey from A to B.
>
> It is at once apparent that this result still holds good if the clock moves from A to B in any polygonal line, and also when the points A and B coincide.
>
> If we assume that the result proved for a polygonal line is also valid for a continuously curved line, we arrive at this result: if one of two synchronous clocks A is moved in a closed curve with constant velocity until it returns to A, the journey lasting t seconds, then by the clock which has remained at rest the travelled clock on its arrival at A will be $(1/2)tv^2/c^2$ second [sic] slow.[28]

In another context, Kopff quoted Einstein from "The Relativity Theory," an article that appeared in *Naturforschende Gesellschaft, Zürich, Vierteljahresschrift*, volume 56, and was published in 1911. Kopff first noted that:

> The consequences deduced from the Lorentz transformation for the passage of clocks apply naturally to the temporal occurrence of arbitrary events. An example which brings out the relations with great clarity is given by Einstein.[29]

According to Kopff, Einstein wrote the following:

> If we placed a living organism in a box, and made it carry through the same to-and-fro motions as the clocks formerly did,

- 65 -

then one could arrange that this organism, after any arbitrary lengthy flight, could be returned to its original spot in a scarcely altered condition, while corresponding organisms which had remained in their original positions had already long since given way to new generations. For the moving organism, the lengthy time of the journey was a mere instant, provided the motion took place with approximately the velocity of light.[30]

But the question then arises why did Einstein also maintain that a train traveling with uniform velocity along an embankment and then experiencing a non-uniform motion for a brief period of time, such as the application of the brakes, could be considered an inertial reference frame throughout its motion, albeit one that experienced a gravitational field at a certain time? Einstein wrote:

> It is certainly true that the observer in the railway carriage experiences a jerk forwards as a result of the application of the brake, and that he recognises in this non-uniformity of motion (retardation) of the carriage. But he is compelled by nobody to refer this jerk to a "real" acceleration (retardation) of the carriage. He might also interpret his experience thus: "My body of reference (the carriage) remains permanently at rest. With reference to it, however, there exists (during the period of application of the brakes) a gravitational field which is directed forwards and which is variable with respect to time. Under the influence of this field, the embankment together with the earth moves non-uniformly in such a manner that their original velocity in the backwards direction is continuously reduced."[31]

This idea is the basis for the principle of equivalence, one statement of which is that inertial and non-inertial reference frames are equivalent with regard to the expression of physical law.[32] In the scenario proposed by Einstein, the train is an inertial reference frame that at some point in time experiences a gravitational field, and this field is associated with the embankment, along with the earth, constituting an accelerating reference frame relative to an observer at rest on the train.

Holton[33] in "Resource Letter SRT-1 on Special Relativity Theory," wrote that Terrell[34] provided "perhaps the best of the recent reviews" of time dilation and clock problems. In his article, Terrell, though arguing that there is indeed only one twin that will experience time dilation, allowed for the idea that

# On the Arbitrary Choice

is the basis for the principle of equivalence, but notes that it is improbable. In his version of the twin paradox, Terrell wrote:

> An alternative which B [the "moving" observer with the clocks at rest in his inertial frame] could choose is to apply the general theory of relativity and assume himself to have been unaccelerated throughout these events. He could, because of the principle of equivalence, account for all observed effects by introducing moving gravitational fields having the nature of plane shock waves, which passed him coincidentally with his application of rocket power (or whatever produced his accelerations as observed by A.) Thus observer B could assume himself to have remained stationary while A and the rest of the universe were accelerated. However this would be a somewhat complicated, coincidental, and physically implausible explanation, although it would agree with the general theory of relativity and would account for all the observations discussed.(35)

It should be noted that there are some fundamental theoretical problems with the conventional solutions proposed to the twin paradox. First, is Einstein's discussion of the railway carriage as the "stationary" frame not implausible? Is it substantially less implausible that an observer on a train sees himself as in the "stationary" frame and the earth as the "moving" frame than Terrell's or anyone else's explanation of the twin paradox? Einstein's idealized example is at the heart of the principle of equivalence. Second, given that the "moving" observer travels for a much longer time in an inertial reference frame relative to the "stationary" observer compared to the time he spends in an accelerating reference frame (during the time the "moving" observer is turning around), how is it that the inertial motion of the "moving" observer over the long period of time gets set, so speak, as really moving by the comparatively very brief acceleration? As this acceleration sets the "moving" twin as really moving, why is there no substantial influence on the time dilation effect which is determined in the large part by the temporal relations for two inertial reference frames moving in a uniform translational manner relative to one another?

     $^2$The essence of the derivation of the Lorentz coordinate transformation equations is that there is a comparison between: 1) the invariant velocity of light in all inertial reference frames, and 2) the magnitude of the uniform translational velocity of two inertial reference frames relative to one another. One of the two

## On the Arbitrary Choice

inertial reference frames in uniform motion relative to one another is designated the "stationary" reference frame, and it is from this reference frame that the other frame is considered "moving" with the uniform translational velocity. This is also the essence of the argument on the relativity of simultaneity, as will be shown. (The designation of the reference frame as "stationary" means essentially that simultaneity is first established in it in the argument on the relativity of simultaneity, even though it might be considered "stationary" because it is the reference frame from which the other reference frame is considered "moving". This is because in Newtonian mechanics, built upon absolute simultaneity, the designation of one or the other of two inertial reference frames in uniform translational motion relative to one another as the "stationary" frame does not have the same significance that it does in the special theory. For example, in the kinematics underlying Newtonian mechanics, spatial lengths of physical existents and the temporal duration of occurrences are the same in both of the inertial frames regardless of which frame is designated the "stationary" reference frame. With regard to these and other essential features of the physical world, in Newtonian mechanics the role of "stationary" and "moving" reference frames can be freely exchanged without affecting the results of measurements. It has been shown that this is not the case in the special theory, and it is the relativity of simultaneity that is responsible for this. Thus, the relativity of simultaneity is indeed the key to designating an inertial reference frame as "stationary".)

[3]There is a difference developing the spatial, as opposed to temporal, relations between inertial reference frames in uniform translational motion relative to one another. In spatial relations, one needs to simultaneously determine the ends of whatever is being measured for spatial length in order to determine this length. If one knows the spatial length of an existent at rest in one inertial frame, one can then follow the world lines of the ends of this existent until they intersect with the time axis of the other inertial frame (which represents simultaneous spatial positions in this reference frame). In the case of temporal relations between these inertial frames, the concern is with measuring temporal durations, that themselves are based on simultaneity in each of the frames, but which, of course do not require that the beginning and end time measurements in a frame be taken simultaneously in that frame. Rather, the concern is with what time a clock at rest in the "stationary" frame will record corresponding to a clock at rest in the "moving" frame.

[4]Einstein noted the reciprocal manner in which space contraction could

be argued. He wrote:

> We envisage a rigid sphere* of radius R, at rest relatively to the moving system k [which is moving relative to K, the stationary system, along the axis x for K and along the  axis for k], and with its centre at the origin of co-ordinates of k....Thus whereas the Y and Z dimensions [that is, along the y and z axes in K] do not appear modified by the motion, the X dimension [that is, the x axis in K] appears shortened in the ratio 1: $\sqrt{(1 - v^2/c^2)}$, i.e., the greater the value of v the greater the shortening....It is clear that the same results hold good of [sic] bodies at rest in the "stationary" system [K], viewed from a system in uniform motion [k].
>
> \* That is, a body possessing spherical form when examined at rest.(36)

[5]One can derive equations 7 and 8 from 5 and 6 and vice versa. But these possibilities of derivation do not negate the point that: 1) in the derivation of equations 5 and 6 one of two inertial reference frames in uniform translational motion relative to one another is designated the "stationary" reference frame and the other inertial reference frame is designated the "moving" reference frame; 2) in the derivation of equations 7 and 8, that reference frame designated "moving" in the derivation of equations 5 and 6 is now the "stationary" reference frame and that reference frame designated "stationary" in the derivation of equations 5 and 6 is now the "moving" reference frame. That is, in either derivation, the time of one inertial reference frame is logically delineated as first, and it is not assumed in the special theory that the time in the other frame will be the same. (That is, the assumption is not made that time is absolute.) It is found that due to the relativity of simultaneity for inertial reference frames in uniform translational motion relative to one another, the time of one reference frame is not the same as that of the other frame.

In Newtonian mechanics, relying on the Galilean coordinate transformation, $t' = t$ so that the derivation of the designation of "stationary" and "moving" inertial frames in the analogous sets of equations is inconsequential. It is inconsequential for the Galilean coordinate transformation itself or for the relation between the temporal durations of an event or the spatial length of a physical existent in the inertial reference frames. That is, in Newtonian mechanics:

$$x' = x - vt \quad (11)$$

and

$$x = x' + vt' \quad (12)$$

As noted, the time relation is $t' = t$. Thus equation 11 transforms to equation 12 simply in the following way. From equation 11,

$$x = x' + vt .$$

Through substitution from $t' = t$,

$$x = x' + vt' .$$

In Newtonian mechanics, clearly the duration of an occurrence in the two inertial reference frames is the same. And it follows from the time relation, $t' = t$, that the length of an object in the two frames is the same. From equation 11, it follows that:

$$\Delta x' = \Delta x - v\Delta t .$$

As $\Delta t' = \Delta t$, if $\Delta t = 0$,

$$\Delta x' = \Delta x$$

Or, using equation 12, one finds in the same manner that:

$$\Delta x = \Delta x' .$$

[6]Another argument concerning the dependence of the Lorentz coordinate transformation equations on the relativity of simultaneity is presented below. It has been shown how spatial length and temporal duration in inertial reference frames in uniform translational motion relative to one another depend on the relativity of simultaneity. The Lorentz coordinate transformation equations can be considered a special case of the equations expressing spatiotemporal relations between the inertial reference frames. One set of spatiotemporal relations between inertial reference frames in uniform translational motion relative to one another is given by:

$$\Delta t = \Delta t'/(1 - v^2/c^2)^{1/2} \quad (2)$$

and

$$\Delta x = \Delta x'(1 - v^2/c^2) \quad (4).$$

Consider the Lorentz coordinate transformation equation:

$$x' = (x - vt)/(1 - v^2/c^2)^{1/2} \quad (7).$$

## On the Arbitrary Choice

x' may be said to stand for $\Delta x'$ in equation 7, specifically where $x_1' = 0$ and $x_2' = x'$. Similarly, x may be said to stand for $\Delta x$ in this equation, where $x_1 = 0$ and $x_2 = x$. Let $t_1$ be the time coordinate corresponding to $x_1$ and $t_2$ be the time coordinate corresponding to $x_2$. If x then does represent a length in W, $t_1 = t_2$. Then equation 7 becomes:

$$x = x'(1 - v^2/c^2)^{1/2} \quad (11)$$

or, in more familiar terms,

$$\Delta x = \Delta x'(1 - v^2/c^2)^{1/2} \quad (4).$$

That it is possible for $t_1 = t_2 = 0$ for any values of $x_1$ and $x_2$, or $\Delta x$, is clear, for $t_1, t_2$, which equal 0, are the time coordinate for the x axis in W. Equation 11 relates space in W and W' as this equation relates the x axis of W and the x' axis of W' where W is considered the "stationary" inertial reference frame and W' is considered the "moving" inertial reference frame.

Similar reasoning can be followed in showing that

$$t' = [t - (v/c^2)x]/(1 - v^2/c^2)^{1/2} \quad (8)$$

can represent a specific form of the relation for time denoted by

$$\Delta t = \Delta t'/(1 - v^2/c^2)^{1/2} \quad (2).$$

Specifically, let t' represent the difference $\Delta t'$ between $t_1' = 0$ and $t_2' = t'$ and t represent the difference $\Delta t$ between $t_1 = 0$ and $t_2 = t$. x can represent the difference $\Delta x$ between $x_1 = 0$ and $x_2 = x$, where $x_1$ is the space coordinate associated with $t_1$ and $x_2$ is the space coordinate associated with $t_2$. In deriving the temporal relation, though, the concern is that the clock is at rest in W' and not at rest in W. That is, W is the "stationary" frame and W' is the "moving" frame in which the clock moving with regard to W is at rest in W'. (In the case of spatial length, even though the rod is at rest in W', nonetheless, in order to determine the length of the rod in W, the ends of the "moving" rod must be determined in W simultaneously, that is simultaneously in accordance with the time in W. This condition is not applicable with regard to time.) Thus, we want $x_1' = x_2'$, where $x_1'$ corresponds to $x_1$ and $x_2'$ corresponds to $x_2$. Then, considering the general derivation from before concerning temporal duration:

$$t_1 = [t_1'(1 - v^2/c^2)^{1/2}] +$$
$$[[v/c^2][x_1'(1 - v^2/c^2)^{1/2} + vt_1]]$$

and

$$t_2 = [t_2'(1 - v^2/c^2)^{1/2}] + [[v/c^2][x_1'(1 - v^2/c^2)^{1/2} + vt_2]] \ .$$

With these two equations, one can derive:

$$t_2 - t_1 = [[t_2'(1 - v^2/c^2)^{1/2}] + [[v/c^2][x_1'(1 - v^2/c^2)^{1/2} + vt_2]]] - [[t_1'(1 - v^2/c^2)^{1/2}] + [[v/c^2][x_1'(1 - v^2/c^2)^{1/2} + vt_1]]]$$

and finally, with arithmetic:

$$t_2 - t_1 = (t_2' - t_1')/(1 - v^2/c^2)^{1/2} \ .$$

Since $t_1 = 0$ and $t_1' = 0$:

$$t = t'/(1 - v^2/c^2)^{1/2} \quad (12)$$

or in more familiar terms,

$$\Delta t = \Delta t'/(1 - v^2/c^2)^{1/2} \quad (2).$$

That it is possible for $x_1' = x_2' = 0$ for any values of $t_1'$ and $t_2'$, or $\Delta t'$, is clear, for $x_1'$, $x_2'$, which equal 0, are the space coordinate for the t' axis in W'. Equation 12 relates time in W and W' as this equation relates the t axis of W and the t' axis of W' where W is considered the "stationary" inertial reference frame and W' is considered the "moving" inertial reference frame.

[7]Furthermore, Einstein himself explicitly acknowledged that c - v is a velocity. As an example, in deriving the Lorentz transformation equations in his original paper on the special theory, Einstein discussed a "stationary" inertial system K and a "moving" inertial system k moving in a uniform translational manner relative to one another. Concerning a ray of light moving in the same direction that k is moving relative to K, Einstein[37] wrote:

> But the ray [of light] moves relatively to the initial point of k [the "moving" system], when measured in the stationary system [K], with the velocity c - v, so that
>
> $$\frac{x'}{c - v} = t \ .$$

## On the Arbitrary Choice

x' is the distance traveled by the ray of light in K, and t is the time taken in K to traverse this distance with the velocity c - v. In another translation of a large section of Einstein's paper, Schwartz[38] translates the textual part of the above quote from Einstein as:

> But, as measured in the stationary system, the ray moves with velocity V - v relative to the origin of k, so that we have
>
> x'/(V - v) = t.
>
> V here is the invariant velocity of light in all inertial reference frames.

It might also be argued that in order to demonstrate the relativity of simultaneity, it is only necessary that the observer considered "stationary" (i.e., the observer in whose frame simultaneity, or the synchronization of clocks, is first established in accordance with one of Einstein's definitions) be aware of the effective velocities c - v and c + v of the lightning flashes relative to the "moving" observer (in the case of the train gedankenexperiment, the observer $O_t$ on the train). If this thesis were correct, there would be an inconsistency in Einstein's arguments because Einstein did not note explicitly any exceptions to the postulate of the special theory of relativity that the velocity of light is invariant in inertial reference frames that are in uniform translational motion relative to one another.

[8]There is, of course, the reversal in the direction of the velocity of the inertial reference frames in the two scenarios for arguing the relativity of simultaneity, and there is also the point that the "moving" reference frame indeed changes spatial coordinates in the "stationary" reference frame over time of the "stationary" reference frame in both scenarios, and thus has velocity in the "stationary" reference frame. The change in direction of the velocity of the inertial reference frames in uniform motion relative to one another is significant because it allows for switching the inertial reference frames with regard to which inertial reference frame will be the "moving" reference frame that is changing spatial coordinates and time coordinates in the "stationary" reference frame and which reference frame will be this "stationary" reference frame. But this change in the direction of the velocity of the inertial reference frames relative to one another is significant with regard to features of the physical world such as spatial length or temporal duration in the special theory, and not in Newtonian mechanics and the kinematics underlying it. It is the relativity of simultaneity that distinguishes the special theory from Newtonian mechanics

## On the Arbitrary Choice

and it is the logical distinction, arbitrarily decided, concerning in which inertial reference frame simultaneity will be established first in the argument that is at the heart of the argument on the relativity of simultaneity.

# On the Arbitrary Choice

# On the Arbitrary Choice

# Appendix 1

## Table I

*Circumstances for Deriving Set 1 and Equations 5 and 6*

| Set 1: $\Delta t' = \Delta t/(1 - v^2/c^2)^{1/2}$ (1); $\Delta x' = \Delta x(1 - v^2/c^2)^{1/2}$ (3) | | |
|---|---|---|
| Equations: $x = (x' + vt')/(1 - v^2/c^2)^{1/2}$ (5); $t = [t' + (v/c^2)x']/(1 - v^2/c^2)^{1/2}$ (6) | | |
| SCENARIO 1: | | |
| Inertial Reference Frame: | W | W' |
| Logical Role in Derivation of Relativity of Simultaneity: | **"Moving" frame.** Time in W' used to determine whether criterion for simultaneity met in W (using same light used to establish simultaneity in W'). | **"Stationary" frame.** Simultaneity established first in this frame. |
| State of Motion of Rod Measured in Equation 3: | Rod measured by observer at rest in W' is at rest in W. The length of this rod at rest in W is longer when measured by an observer at rest in W than when measured by an observer at rest in W'. This rod that is measured and that is of identical construction to the rod at rest in W' that is used to measure it in W' is in Scenario 2, instead, used to measure the length of the latter rod (the rod at rest in W') where the latter rod is the "moving" rod. | Rod measured by observer at rest in W' is "moving" relative to W'. $\Delta x'$ is determined for "moving" measuring rod. Need to determine positions of ends of rod in W' simultaneously. To measure $\Delta x'$, observer at rest in W' uses rod of identical construction to the rod being measured. This rod used to measure $\Delta x'$ is at rest in W'. The length of the rod "moving" in W' is shorter when measured by an observer at rest in W' than when measured by an observer at rest in W. |

# On the Arbitrary Choice

## Table I (continued)

| | SCENARIO 1: | |
|---|---|---|
| Inertial Reference Frame: | W | W' |
| State of Motion of Clocks Measured in Equation 1: | Clock (or clocks) measured by observer at rest in W' is at rest in W.<br><br>The rate of this clock (or clocks) at rest in W is faster when measured by an observer at rest in W than when measured by an observer at rest in W'.<br><br>This clock (or clocks) that is measured and that is (or are) of identical construction to the clocks at rest in W' that are used to measure the clock (or clocks) at rest in W is in Scenario 2, instead, used to measure the rate of the latter clocks (the clocks at rest in W') where the latter clocks are the "moving" clocks. | Clock (or clocks) measured by observer at rest in W' is "moving" relative to W'.<br><br>Clocks at rest in W' are used to determine $\Delta t'$ corresponding to $\Delta t$ in W. Observers at rest in W' use clocks of identical construction to that (or those) clock (or clocks) at rest in W that is (or are) being measured.<br><br>The rate of the clock (or clocks) "moving" in W' is (or are) slower when measured by an observer at rest in W' than when measured by an observer at rest in W. |
| State of Motion of Reference Frame for Observer at Rest in that Frame: | "Stationary" for observer at rest in W. | "Stationary" for observer at rest in W'. |
| State of Motion of Reference Frame for Observer at Rest in Other Frame: | "Moving" for observer at rest in W'. | "Moving" for observer at rest in W. |

# On the Arbitrary Choice

## Table II

*Circumstances for Deriving Set 2 and Equations 7 and 8*

| Set 2: $\Delta t = \Delta t'/(1 - v^2/c^2)^{1/2}$ (2); $\Delta x = \Delta x'(1 - v^2/c^2)^{1/2}$ (4) | | |
|---|---|---|
| Equations: $x' = (x - vt)/(1 - v^2/c^2)^{1/2}$ (7); $t' = [t - (v/c^2)x]/(1 - v^2/c^2)^{1/2}$ (8) | | |
| SCENARIO 2: | | |
| Inertial Reference Frame: | W | W' |
| Logical Role in Derivation of Relativity of Simultaneity: | **"Stationary" frame.** Simultaneity established first in this frame. | **"Moving" frame.** Time in W used to determine whether criterion for simultaneity met in W' (using same light used to establish simultaneity in W). |
| State of Motion of Rod Measured in Equation 4: | Rod measured by observer at rest in W is "moving" relative to W. $\Delta x$ is determined for "moving" measuring rod. Need to determine positions of ends of rod in W simultaneously. To measure $\Delta x$, observer at rest in W uses rod of identical construction to the rod being measured. This rod used to measure $\Delta x$ is at rest in W. The length of the rod "moving" in W is shorter when measured by an observer at rest in W than when measured by an observer at rest in W'. | Rod measured by observer at rest in W is at rest in W'. The length of this rod at rest in W' is longer when measured by an observer at rest in W' than when measured by an observer at rest in W. This rod that is measured and that is of identical construction to the rod at rest in W that is used to measure it in W is in Scenario 1, instead, used to measure the length of the latter rod (the rod at rest in W) where the latter rod is the "moving" rod. |

Table II (continued)

| SCENARIO 2: | | |
|---|---|---|
| Inertial Reference Frame: | W | W' |
| State of Motion of Clocks Measured in Equation 2: | Clock (or clocks) measured by observer at rest in W' is "moving" relative to W'.<br><br>Clocks at rest in W are used to determine $\Delta t$ corresponding to $\Delta t'$ in W'. Observers at rest in W' use clocks of identical construction to that (or those) clock (or clocks) at rest in W' that is (or are) being measured.<br><br>The rate of the clock (or clocks) "moving" in W is (or are) slower when measured by an observer at rest in W than when measured by an observer at rest in W'. | Clock (or clocks) measured by observer at rest in W is at rest in W'.<br><br>The rate of this clock (or clocks) at rest in W' is faster when measured by an observer at rest in W' than when measured by an observer at rest in W.<br><br>This clock (or clocks) that is measured and that is (or are) of identical construction to the clocks at rest in W that are used to measure the clock (or clocks) at rest in W' is in Scenario 1, instead, used to measure the rate of the latter clocks (the clocks at rest in W) where the latter clocks are the "moving" clocks. |
| State of Motion of Reference Frame for Observer at Rest in that Frame: | "Stationary" for observer at rest in W. | "Stationary" for observer at rest in W'. |
| State of Motion of Reference Frame for Observer at Rest in Other Frame: | "Moving" for observer at rest in W'. | "Moving" for observer at rest in W. |